\documentclass[aps,pra,amsmath,amssymb,floatfix,twocolumn,amsmath,superscriptaddress,twocolumn,nofootinbib,tighten,letterpaper]{revtex4-2}
\usepackage[colorlinks,linkcolor=blue,citecolor=blue,urlcolor=blue]{hyperref}
\usepackage{multirow}
\usepackage{subfigure}
\usepackage{color}
\usepackage{mathrsfs}
\usepackage{hyperref}
\usepackage[normalem]{ulem}
\usepackage{bm}

\usepackage{amssymb}
\usepackage{amsmath}
\renewcommand\vec[1]{\ensuremath\boldsymbol{#1}}

\usepackage{tabularray}

\usepackage{array} 
\newcolumntype{P}[1]{>{\centering\arraybackslash}p{#1}}

\definecolor{RowColor}{rgb}{0.88,1,0.9}

\usepackage{amsfonts, relsize, color, mathtools, physics}
\usepackage{graphics}
\usepackage{graphicx}
\usepackage{subfigure}
\usepackage{color}
\usepackage{comment}

\usepackage{tikz}
\usetikzlibrary{shapes.geometric}

\definecolor{markergreen}{HTML}{008000}

\newcommand{\markercircle}{\tikz[baseline=-0.6ex]\fill[markergreen] (0,0) circle (0.6ex);}
\newcommand{\markersquare}{\tikz[baseline=-0.6ex]\fill[markergreen] (-0.55ex,-0.55ex) rectangle (0.55ex,0.55ex);}
\newcommand{\markertriangle}{\tikz[baseline=-0.6ex]\fill[markergreen] (90:0.8ex) -- (210:0.8ex) -- (330:0.8ex) -- cycle;}
\newcommand{\markerinvtriangle}{\tikz[baseline=-0.6ex]\fill[markergreen] (270:0.8ex) -- (30:0.8ex) -- (150:0.8ex) -- cycle;}
\newcommand{\markerstar}{\tikz[baseline=-0.6ex]\node[star, star points=5, star point ratio=2.25,
    fill=markergreen, inner sep=0pt, minimum size=1.9ex] at (0,0) {};}
    
\definecolor{markermagenta}{HTML}{FF00FF}

\newcommand{\markerpentagon}[1][markergreen]{\tikz[baseline=-0.6ex]\fill[#1]
    (90:0.75ex) -- (162:0.75ex) -- (234:0.75ex) -- (306:0.75ex) -- (18:0.75ex) -- cycle;}

\begin{document}

\title{Topological insulators on site-eliminated, projected, and patterned fractal lattices}

\author{Aiden J.\ Mains}~\thanks{Equal contributors}
\affiliation{Department of Physics, Lehigh University, Bethlehem, Pennsylvania 18015, USA}

\author{Daniel J.\ Salib}~\thanks{Equal contributors}
\affiliation{Department of Physics, Lehigh University, Bethlehem, Pennsylvania 18015, USA}

\author{Christopher A.\ Leong}
\affiliation{Department of Physics, Lehigh University, Bethlehem, Pennsylvania 18015, USA}

\author{Bitan Roy}~\thanks{Corresponding author: bitan.roy@lehigh.edu}
\affiliation{Department of Physics, Lehigh University, Bethlehem, Pennsylvania 18015, USA}

\date{\today}

\begin{abstract}
Identifying fractal lattices of fractional (Hausdorff) dimension as geometric descendants of parent crystals, we showcase the successful realization of canonical topological insulators in one-, two-, and three-dimensional crystals on their self-similar descents using three complementary approaches, starting from lattice-regularized models in integer dimensions. (a) In the first approach, we construct the fractal Hamiltonian by removing all parent-lattice sites that do not belong to such a network (method of site elimination). (b) In the second one, we integrate or project out the same set of sites to obtain an effective Hamiltonian for the fractal lattice (method of projection or renormalization). (c) In the third approach, we retain the parent-crystal lattice intact but spatially pattern the on-site mass parameter, which controls the topology of the parent insulating system, such that it assumes two topologically distinct values on the fractal network and its complement (patterned fractal lattices). By numerically diagonalizing the resulting Hamiltonians and computing appropriate topological invariants or local markers, corroborating the bulk-boundary correspondence through gapless modes at outer or inner boundaries, we establish Su-Schrieffer-Heeger topology on the Cantor set using all three approaches, Haldane topology on the hexaflake [using (a) and (b)] and Qi-Wu-Zhang topology on the Sierpi\'nski carpet [using (c)], and strong $Z_2$ topological insulating phases on the tiled Menger sponge from methods (a), (b), and (c). These findings complement and substantially extend the previously reported realization of Qi-Wu-Zhang topology on the Sierpi\'nski carpet using methods (a) and (b) [\href{https://journals.aps.org/prb/abstract/10.1103/PhysRevB.110.L241302}{Phys.\ Rev.\ B {\bf 110}, L241302 (2024)}]. 
\end{abstract}

\maketitle

\section{Introduction}

Translational symmetry gives rise to Bloch states as solutions of the Schr\"odinger equation in quantum crystals. When endowed with nontrivial topology and geometry over the Brillouin zone, these states give rise to topological crystals. The associated bulk topological invariants manifest through gapless boundary modes at the interface between the topological crystal and vacuum, embodying the celebrated bulk-boundary correspondence~\cite{TopoGen:1, TopoGen:2, TopoGen:3, TopoGen:4, TopoGen:5, TopoGen:6, TopoGen:7, TopoGen:8, TopoGen:9, TopoGen:10, TopoGen:11, TopoGen:12}. Translational symmetry further divides topological phases into two broad categories, depending on whether the requisite band inversion occurs at the center of the Brillouin zone (the $\Gamma$ point) or at finite momentum, giving rise to translationally inert and active topological phases, respectively. Incorporating discrete rotational symmetries into this classification further enriches the landscape through crystalline topological phases, characterized by simultaneous band inversions at multiple high-symmetry points in the Brillouin zone related by discrete rotations. The classification and characterization of these topological phases have remained central themes in condensed matter physics over the past few decades, more recently culminating in the development of topological quantum chemistry, as a framework for identifying them in real materials~\cite{TopCrys:1, TopCrys:2, TopCrys:3, TopCrys:4, TopCrys:5, TopCrys:6, TopCrys:7}.

Nature, however, hosts a rich variety of structures beyond the realm of perfect crystalline order, including fractals, quasicrystals, and amorphous lattices, which lack some or all of the translational and rotational symmetries of crystals. Nevertheless, such noncrystalline arrangements of atoms or sites can often be viewed as geometric descendants of parent crystals. A regular fractal lattice, for instance, can be constructed by systematically removing selected sites from a parent crystal, thereby replacing translational symmetry with self-similarity while preserving the overall rotational symmetry of the parent lattice~\cite{FractalGen:1}. A quasicrystal, in contrast, arises from projecting a higher-dimensional parent crystal onto a hyperplane embedded within it at an irrational orientation. Such aperiodic structures lack translational symmetry but display quasi-long-range order, manifested in their X-ray diffraction patterns, and can exhibit emergent rotational symmetries otherwise forbidden in conventional crystals. Prominent examples include the five-fold and eight-fold rotational symmetries of Penrose and Ammann--Beenker tilings, respectively. Closely related are rational approximants of quasicrystals, obtained by projecting the parent crystal onto a hyperplane at a rational orientation. These structures exhibit emergent periodicity in lower dimensions, distinct from that of the original parent crystal~\cite{QCGen:1,  QCGen:2,  QCGen:3,  QCGen:4}. Finally, amorphous lattices can be constructed by randomly selecting a finite fraction of sites from an otherwise ordered parent crystal, thereby eliminating both translational and rotational symmetries~\cite{AmorphousGen:1}. Within this framework, irregular fractal lattices and amorphous systems with partial orders can also be considered as geometric descendants of parent crystals, which partially retain some of its symmetries or orders.

\begin{figure}[t]
    \includegraphics[width=1.0\linewidth]{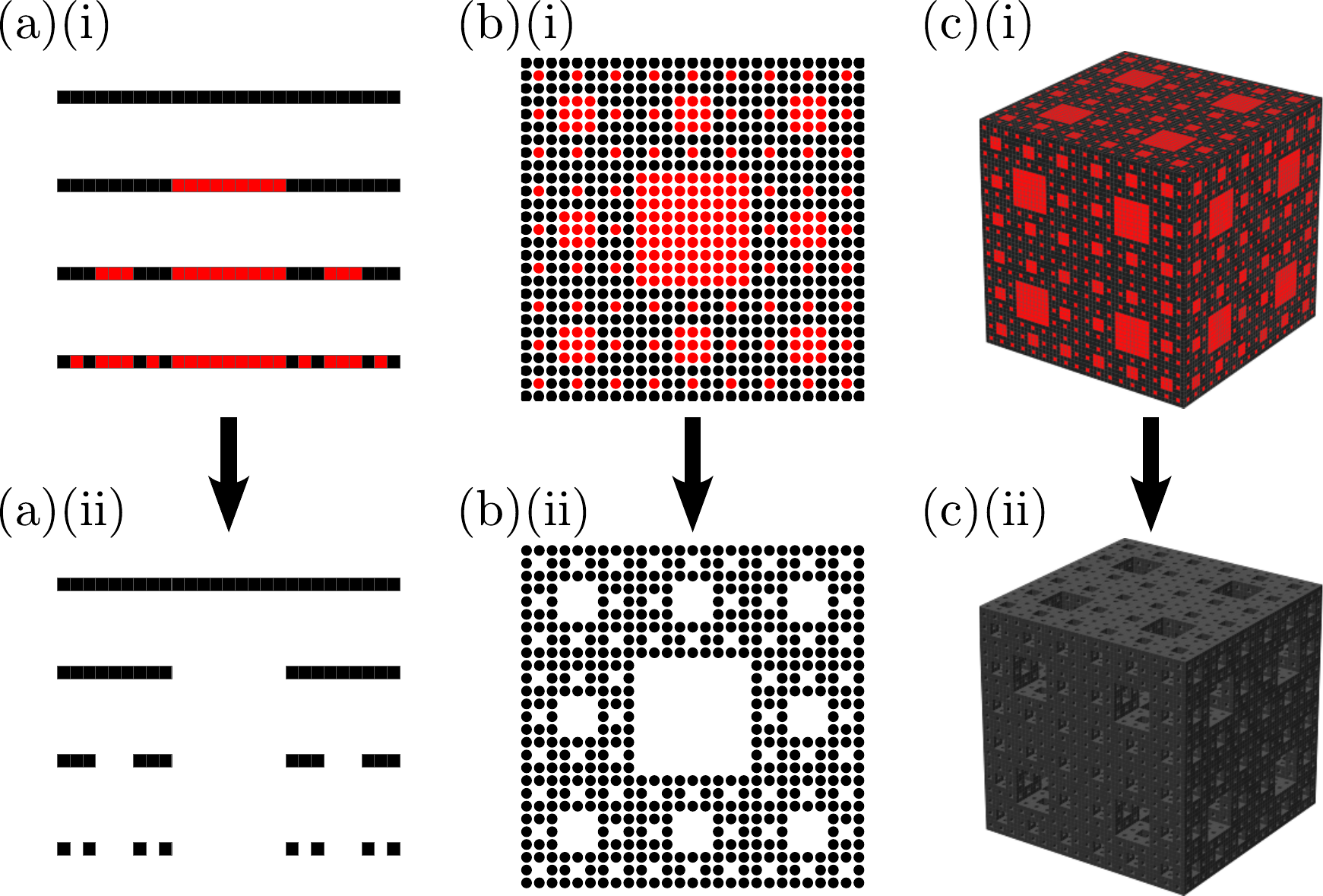}
    \caption{Three fractal textures studied in this work: (a) Cantor set, (b) Sierpi\'nski carpet, and (c) tiled Menger sponge, constructed by arranging eight blocks of conventional Menger sponge in a $2 \times 2 \times 2$ configuration. For each texture, the top panel shows the full parent lattice: (a) a one-dimensional chain, (b) a square lattice, and (c) a cubic lattice. In the substituted-mass construction [(i)], the black and red sites carry on-site masses $M_0$ and $M_{\rm alt}$, respectively, occupying the sites belonging to the resulting fractal lattice [shown in (ii)] and its complement. Together, these configurations constitute the family of patterned fractal lattices. The bottom panel shows the corresponding fractal lattices obtained by removing (method of site elimination) or projecting out (method of projection or renormalization) all the red sites from the parent lattices, resulting in a Hamiltonian where only the fractal lattice sites are operative.
    }~\label{fig:hypercubic_illustration}
\end{figure}

The structural diversity of such material platforms, together with their underlying connection to parent crystals established above, raises a fascinating question: can the rich catalog of crystal-based topological phases be extended to fractal lattices, quasicrystals, and amorphous materials? Beyond its fundamental significance, an affirmative answer could have far-reaching technological implications, potentially enabling structurally imperfect materials to host topologically protected, fault-tolerant responses, with promising applications in low-power logic devices, for example.

Although recent years have witnessed growing interest in this direction, existing studies have largely focused on isolated, model-specific realizations of topological phases in selected noncrystalline structures, including fractal lattices~\cite{frac:th1, frac:th2, frac:th3, frac:th4, frac:th5, frac:th6, frac:th7, frac:th8, frac:th9, frac:th10, frac:th11, frac:th12, frac:th13, frac:th14, frac:th15, frac:th16, frac:th17, frac:th18, frac:th19}, quasicrystals~\cite{QCtop:1, QCtop:2, QCtop:3, QCtop:4, QCtop:5}, and amorphous materials~\cite{amorphous:1, amorphous:2, amorphous:3, amorphous:4, amorphous:5, amorphous:6, amorphous:7, amorphous:8, amorphous:9, amorphous:10, amorphous:11, amorphous:12, amorphous:13}. Developing a unified theoretical framework to harness parent-crystal topology across fractional (Hausdorff) dimensions (in fractal lattices), lower-dimensional systems (in quasicrystals and their rational crystalline approximants), and structurally disordered systems without a well-defined dimensionality (in amorphous lattices), under the overarching theme of \emph{topology beyond crystalline order}, remains in its infancy~\cite{GenConstruct:1, GenConstruct:2, GenConstruct:3, GenConstruct:4}. The present work takes a critical step toward this goal by demonstrating the realization of crystal-based canonical topological models in one, two, and three dimensions on their geometric descendant fractal lattices using two complementary theoretical approaches, and by introducing the concept of \emph{patterned topological fractals}.

\subsection{Methodology}~\label{subsec:methods}

\begin{figure}[t!]
    \centering
    \includegraphics[width=1.00\linewidth]{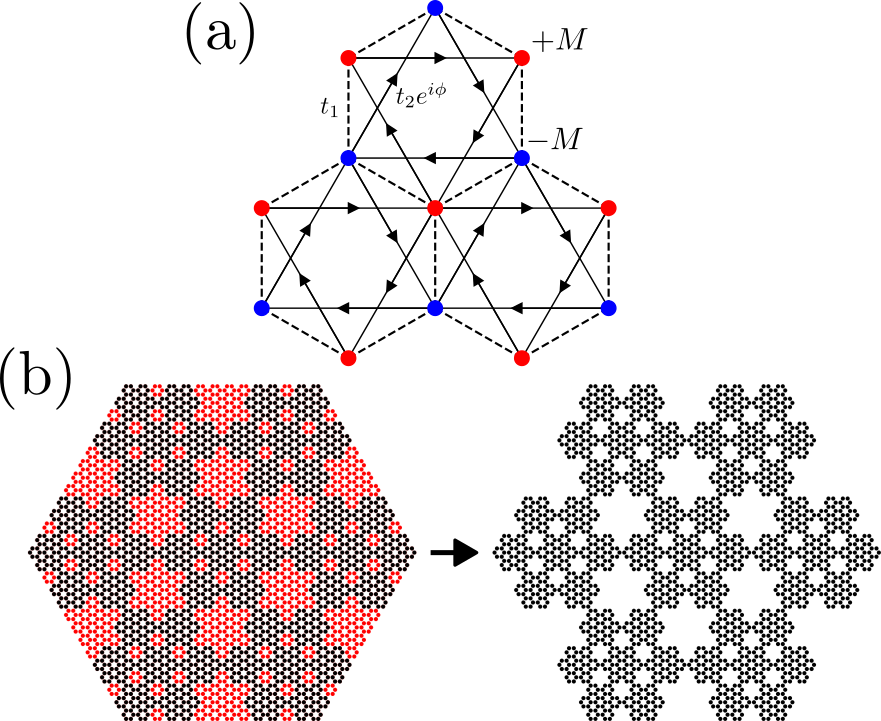}
    \caption{(a) Illustration of the Haldane model with a staggered on-site Semenoff mass ($\pm M$ on the sites of two complementary sublattices of the honeycomb lattice), real nearest-neighbor hopping amplitude $t_1$, yielding massless Dirac fermions, and a complex next-nearest-neighbor hopping $t_2 e^{i \phi}$ which breaks the time-reversal symmetry and produces a Chern insulator. (b) Construction of a third generation hexaflake fractal lattice with $2058$ sites from its parent honeycomb lattice of $3282$ sites. Red sites are removed (method of site elimination) or integrated out (methods of projection or renormalization) resulting in a Hamiltonian where only the fractal-lattice sites remain operative.}
    \label{fig:haldane_setup_figure}
\end{figure}

The discussion is best preceded by a pedagogical overview of the three methodologies employed here to construct Hamiltonians on fractal networks. These constructions rely on two key observations. (1) A fractal lattice can be identified within its parent crystal by specifying the set of sites to be eliminated or integrated out or patterned (as detailed below) to generate the desired self-similar network. In Figs.~\ref{fig:hypercubic_illustration} and~\ref{fig:haldane_setup_figure}, the sites belonging to the fractal lattices are shown in black, while the remaining sites of the parent lattices are shown in red. (2) Canonical lattice-regularized models of topological insulators across spatial dimensions and symmetry classes can be formulated as effective single-particle massive Dirac Hamiltonians, which we denote by $H_{\rm parent}$. Any such quadratic Hamiltonian on the parent lattice can be expressed in the following block-matrix form
\begin{equation}~\label{eq:blockmatrix}
H_{\rm parent} = \left( \begin{array}{cc}
H_{\bullet \bullet} & H_{\bullet {\color{red} \bullet}} \\
H_{{\color{red} \bullet} \bullet} & H_{{\color{red}\bullet \bullet}}
\end{array}
\right),
\end{equation}
where $H_{\bullet\bullet}$ and $H_{{\color{red}\bullet\bullet}}$ act exclusively on the black and red sites, respectively, while the off-diagonal blocks, $H_{\bullet {\color{red} \bullet}}$ and $H_{{\color{red} \bullet} \bullet}$, describe couplings between the two sectors.

In the first approach, termed the method of site elimination and hereafter denoted by (a), we simply remove the red sites, without accounting for their coupling to the black sites. The resulting fractal-lattice Hamiltonian is
\begin{equation}\label{eq:siteeliminated}
H_{\rm SE}=H_{\bullet\bullet}\;.
\end{equation}
In the second approach, termed the method of projection or renormalization and denoted by (b), we systematically integrate out the red sites within the path-integral formalism, yielding the effective (renormalized) Hamiltonian for the fractal lattice
\begin{equation}\label{eq:renormalized}
H_{\rm R}=H_{\bullet\bullet}
-H_{\bullet {\color{red}\bullet}}\,
H^{-1}_{{\color{red}\bullet\bullet}}\,
H_{{\color{red}\bullet}\bullet},
\end{equation}
provided that $H^{-1}_{{\color{red}\bullet\bullet}}$ exists. This condition can be satisfied as possible singularities (zero modes of $H_{{\color{red}\bullet\bullet}}$) are always isolated, and therefore can be regularized by taking a proper limiting procedure~\cite{silvester}. The possible isolated singularities can, for example, be regularized, by taking $H_{{\color{red}\bullet\bullet}} \to H_{\color{red}\bullet\bullet}-\epsilon$ with a small $\epsilon$. The expression for $H_{\rm R}$ is the Schur complement of $H_{\rm parent}$~\cite{GenConstruct:1, GenConstruct:2, GenConstruct:3, GenConstruct:4, GenConstruct:5, GenConstruct:6, GenConstruct:7}. Both $H_{\rm SE}$ and $H_{\rm R}$ act on the same reduced Hilbert space, whose dimensionality is smaller than that of the parent system.

In the third approach, termed the method of patterned fractal lattices and denoted by (c), no sites are removed or integrated out. Instead, the parent lattice is retained intact, while its on-site mass parameter is spatially patterned to assume two topologically distinct values, $M_0$ on the fractal network and $M_{\rm alt}$ on its complement. Accordingly, the Hamiltonian of patterned fractal lattices is $H_{\rm pattern}=H_{\rm parent}$, but with two distinct mass parameters: $M_0$ on the black sites entering $H_{\bullet\bullet}$ and $M_{\rm alt}$ on the red sites entering $H_{{\color{red}\bullet\bullet}}$. The off-diagonal blocks, $H_{\bullet {\color{red} \bullet}}$ and $H_{{\color{red} \bullet} \bullet}$, remain unaffected by this mass patterning, as they encode only intersite hopping processes. Using these three complementary approaches, we demonstrate the emergence of topological insulating phases on fractal lattices embedded in one, two, and three spatial dimensions, as detailed in the following key results.

\subsection{Summary of main results}

\begin{figure*}[t!]
    \centering
    \includegraphics[width=\linewidth]{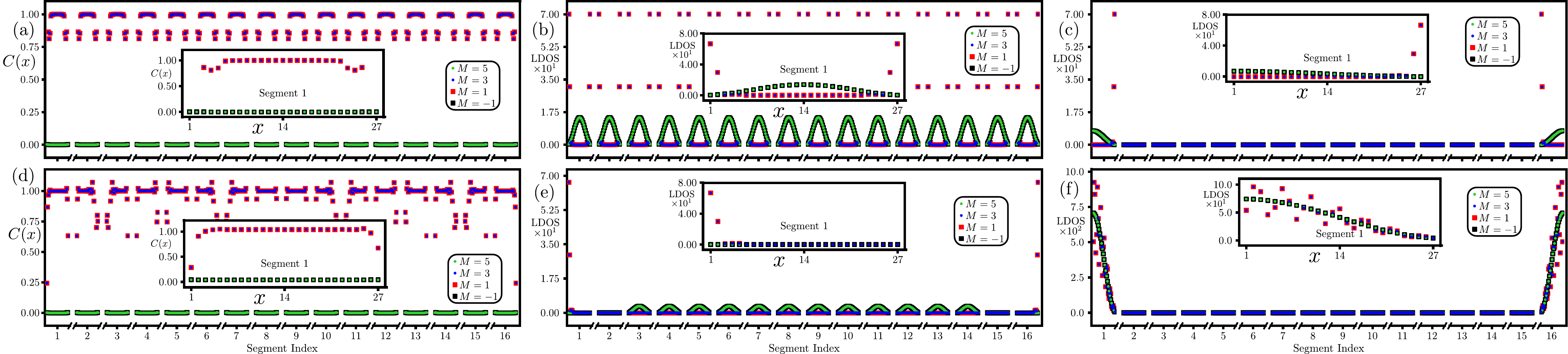}
    \caption{Local topological marker $C(x)$ [(a) and (d)], defined in Eq.~\eqref{eq:utm_1d}, and the joint local density of states for the two particle-hole symmetric closest to zero-energy modes with open [(b) and (e)] and periodic [(c) and (f)] boundary conditions at every site at $x$ of the fourth generation ($g=4$) one-dimensional (1D) Cantor set containing $L=2187$ sites [Fig.~\ref{fig:hypercubic_illustration}(a,ii)]. The results are obtained via method of site elimination [(a)-(c)] and method of projection or renormalization [(d)-(f)] for various choices of the on-site mass parameter $M$ (see legends) upon setting $t=B=1$ in the parent Su–Schrieffer–Heeger model on a 1D chain [Eqs.~\eqref{eq:ham_1d} and~\eqref{eq:ham_1d_TB}]. In each panel, the inset shows the first $27$ sites along the chain. Note that with periodic boundary conditions, the leftmost and rightmost segments of the Cantor set are glued to form a single segment with twice the length. Consequently, the local density of states of such modes shows signatures of (c) boundary localization (method of site elimination) or (f) dominant bulk localization (method of projection of renormalization) on such a segment even with periodic boundary conditions. Results with method of projection or renomalization are obtained by taking $H_{{\color{red}\bullet\bullet}} \to H_{\color{red}\bullet\bullet}-\epsilon$ in Eq.~\eqref{eq:renormalized} with $\epsilon=0.01$ to regulate isolated singularities in the spectrum of $H^{-1}_{{\color{red}\bullet\bullet}}$. See Sec.~\ref{sec:topofrac} for discussion.
    }~\label{fig:cantor_marker}
\end{figure*}

In one dimension, we consider a parent 1D chain described by the Su-Schrieffer-Heeger (SSH) model~\cite{SSH:1, SSH:2, SSH:3} and investigate its topological fingerprints on its fractal descendant, the Cantor set, characterized by the Hausdorff dimension $d_{\rm H}=\log_{3}2 \approx 0.631$. Using all three construction methods, we demonstrate the realization of 1D topological insulating phases on the Cantor set. For methods (a) and (b), the topological phase is characterized by a local topological marker that assumes quantized values over the majority of sites on each segment of the Cantor set, whereas the patterned Cantor set [method (c)] is characterized by a quantized polarization, a global topological invariant. In all three approaches, the topological phase supports near-zero-energy modes localized at boundaries with vacuum [methods (a), (b), and (c)] or at interfaces between regions with distinct on-site mass parameters $M$ and $M_{\rm alt}$ [method (c)], with their precise locations depending on the boundary conditions (open or periodic). All three methods also support a trivial insulating phase, characterized by a vanishing local topological marker or polarization. Interestingly, near-zero-energy modes also emerge in this trivial phase for method (c), where they are strongly localized at the interfaces between regions with distinct mass parameters $M$ and $M_{\rm alt}$ within the interior of the patterned Cantor set. Further details of this model are presented in Sec.~\ref{subsec:1D}, with the results summarized in Fig.~\ref{fig:cantor_marker} [methods (a) and (b)] and Fig.~\ref{fig:cantor_phase_diagram} [method (c)].

For two-dimensional systems, we recently demonstrated the realization of translationally inert, translationally active, and crystalline topological insulators of a generalized square-lattice Qi-Wu-Zhang (QWZ) model for quantum anomalous Hall insulators (QAHIs) on its geometric descendant, the Sierpi\'nski carpet, with the Hausdorff dimension $d_{\rm H}=\log_{3}8 \approx 1.893$, using methods (a) and (b)~\cite{QWZ, GenConstruct:3}. To further explore the capacity of the Sierpi\'nski carpet to inherit topological phases from its parent square lattice, here we revisit the same model (within a slightly restricted parameter space) using method (c). In parallel, to extend the applicability of methods (a) and (b) to two-dimensional topological fractals, we investigate the honeycomb lattice-based Haldane model for QAHIs~\cite{haldane1988, semenoff1984} on its geometric descendant hexaflake fractal lattices, with $d_{\rm H}=\log_{3}7 \approx 1.771$. In both setups, we realize time-reversal-symmetry-breaking topological insulating phases, characterized by a quantized Bott index (a global topological invariant) and corroborated by the bulk-boundary correspondence through near-zero-energy modes. Their localization properties are qualitatively similar to those found earlier for the 1D Cantor set. Further details are provided in Sec.~\ref{subsec:2D}, with the results presented in Figs.~\ref{fig:haldane_phase_diagram} and~\ref{fig:haldane_ldos} for the Haldane model on the hexaflake, and Fig.~\ref{fig:carpet_phase_diagram} for the QWZ model on the Sierpi\'nski carpet. In all these cases as well, there exists extended parameter regimes over which the system describes a trivial insulator with a zero Bott index.

In three dimensions, we consider a tiled Menger sponge constructed by arranging eight conventional Menger sponges, each with a Hausdorff dimension of $d_{\rm H}=\log_{3}20\approx 2.729$, in a $2\times2\times2$ configuration such that the center of the resulting structure is not hollow. Since the tiled structure comprises a finite union of Menger sponges, its Hausdorff dimension remains $d_{\rm H}=\log_{3}20\approx 2.729$, although the system then loses global self-similarity. Then, starting from the canonical model of a strong $Z_2$ topological insulator on the parent cubic lattice in terms of three-dimensional analog of the Bernevig-Hughes-Zhang (BHZ) model~\cite{TopoGen:8, TopoGen:12}, we demonstrate its realizations on the tiled Menger sponge using all three construction methods, (a), (b), and (c). The resulting topological phases exhibit a quantized local topological marker over the majority of bulk sites and satisfy the bulk-boundary correspondence through near-zero-energy modes, whose localization properties are qualitatively similar to those found in the 1D and 2D topological fractals. These results are discussed in Sec.~\ref{subsec:3D} and presented in Figs.~\ref{fig:sponge_marker} and~\ref{fig:sponge_substituted_marker}. Over an extended parameter regime, the tiled Menger sponge also supports normal insulating phases, characterized by a vanishing local topological marker throughout the system for methods (a) and (b), and by its nonquantized values for method (c).

Finally, we investigate the robustness of fractal topological insulators against random pointlike charge impurities. Since local topological markers fluctuate even in clean systems, with these fluctuations becoming more pronounced in the presence of randomness in the system, they are not suitable for identifying disorder-driven transitions between topological and trivial insulating phases. We, therefore, exclusively focus on the Haldane model on hexaflake fractals and use the disorder-averaged Bott index to demonstrate the stability of the resulting QAHIs against sufficiently weak disorder using both methods (a) and (b). For sufficiently strong disorder, the system invariably transitions into a trivial insulating phase, characterized by a vanishing disorder-averaged Bott index. Details of this analysis are presented in Sec.~\ref{sec:disorder}, with the results shown in Fig.~\ref{fig:haldane_disorder}.

\subsection{Organization}

The rest of the manuscript is organized as follows. Section~\ref{sec:topofrac} is devoted to the discussion on the realization of various crystal-based canonical topological insulators on their geometric descendant fractal lattices in one dimension [Sec.~\ref{subsec:1D}], two dimensions [Sec.~\ref{subsec:2D}], and three dimensions [Sec.~\ref{subsec:3D}]. The stability of Haldane model-based QAHIs on hexaflake fractal lattices against random charge impurities is discussed in Sec.~\ref{sec:disorder}. We summarize the findings, discuss possible future directions, and highlight promising material platforms to test our predictions in Sec.~\ref{sec:summary}.

\section{Topological insulators on fractal lattices}~\label{sec:topofrac}

We now demonstrate realizations of topological insulators on fractal lattices in one, two, and three dimensions in three successive subsections. In each case, we begin with the canonical lattice-regularized model(s) of the corresponding topological insulator(s) and construct the Hamiltonian on the associated geometric-descendant fractal lattice(s) using the three different methods detailed in Sec.~\ref{subsec:methods}. We then numerically diagonalize the resulting fractal Hamiltonians to identify signatures of topological insulating phases in these self-similar networks. In particular, we compute the appropriate local topological markers or global topological invariant and demonstrate their consistency with the iconic bulk-boundary correspondence through the emergence of near-zero-energy modes. The localization properties of such modes with open boundary conditions (OBCs) and periodic boundary conditions (PBCs), in turn, reveal fascinating features of the underlying fractal geometry.

\subsection{One dimension}~\label{subsec:1D}

We begin with the SSH model of spinless fermions in one dimension, whose Bloch Hamiltonian takes the form
\begin{eqnarray}~\label{eq:ham_1d}
    H^{\rm SSH}_{\rm 1D}(\vec{k})=t\sin(k_x a) \tau_1 + \left[ M-2B \left\{ 1-\cos(k_xa) \right\} \right]\tau_2,
\end{eqnarray}
where $a$ is the lattice spacing of the underlying 1D chain, $\{ \tau_\mu \}$ denotes the set of Pauli matrices acting on the sublattice or orbital degrees of freedom, and $k_x$ is the momentum along the $x$ direction. The model preserves three symmetries: (1) time-reversal symmetry, generated by ${\mathcal T}=\tau_3 {\mathcal K}$, where ${\mathcal K}$ denotes complex conjugation; (2) particle-hole symmetry, generated by ${\mathcal C}={\mathcal K}$; and (3) sublattice symmetry, generated by ${\mathcal S}=\tau_3$. Accordingly, the Hamiltonian in Eq.~\eqref{eq:ham_1d} belongs to symmetry class BDI of the tenfold classification scheme and realizes a topological insulating phase for $0< M/B <4$. For all other values of $M/B$, the insulating phase is topologically trivial. The Zak phase is quantized to $+1$ in the topological phase and vanishes in the trivial phase~\cite{ZakPhase}. Throughout, we set $t=B=1$ for all the numerical calculations.

On a 1D chain, the tight-binding Hamiltonian corresponding to $H^{\rm SSH}_{\rm 1D}(\vec{k})$ [Eq.~\eqref{eq:ham_1d}] is readily obtained via Fourier transformation and takes the exact form
\begin{eqnarray}~\label{eq:ham_1d_TB}
H^{\rm SSH}_{\rm 1D, TB} &=& \sum_{x} \bigg\{ \left( \Psi^\dagger_{x} \left[ \frac{t}{2 i} \; \tau_1 + B \tau_2 \right]\Psi_{x+\hat{e}_x} + {\rm H.c.} \right) \nonumber \\
&+& \Psi^\dagger_x \left[ (M+2 B) \tau_2 \right]\Psi_x \bigg\} + \cdots.
\end{eqnarray}
Here, $\Psi^\top_x=\left( c_{x,+}, c_{x,-} \right)$ with $c_{x,\tau}$ denoting the fermionic annihilation operator on a site at $x$ in the orbital with parity eigenvalue $\tau=\pm$, H.c.\ denotes the Hermitian conjugate, and $\hat{e}_x$ is the unit vector along the $x$ direction. For brevity, we omit the additional hopping terms required to impose the PBCs, as indicated by the ellipsis. The above tight-binding Hamiltonian on the parent 1D chain can then be recast in the block-matrix form of Eq.~\eqref{eq:blockmatrix}, with $H_{\rm parent} \equiv H^{\rm SSH}_{\rm 1D, TB}$, from which we construct the Hamiltonian for the Cantor set fractals as follows.

\begin{figure*}[t]
    \centering
    \includegraphics[width=1.00\linewidth]{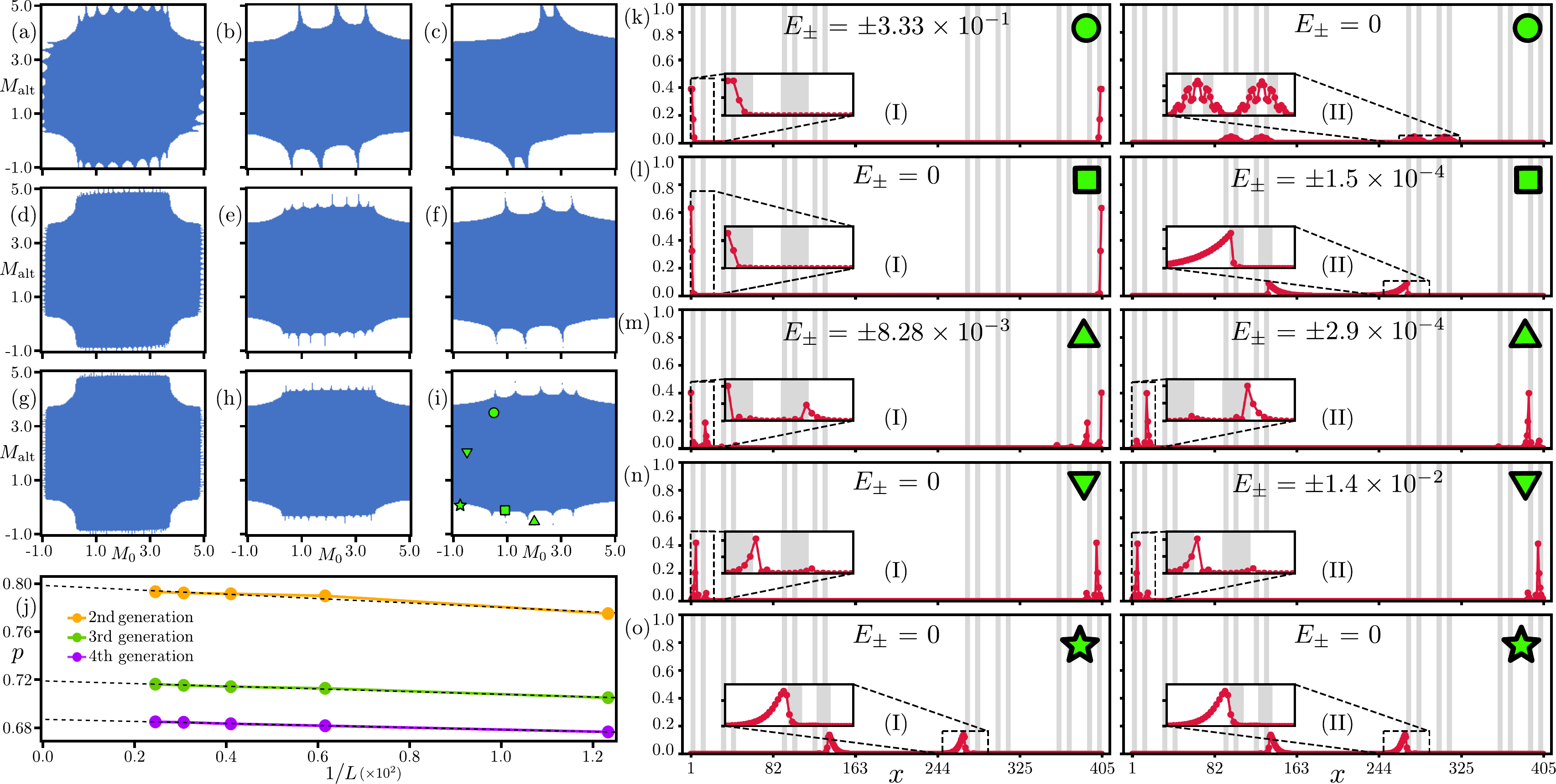}
    \caption{Phase diagram [(a)-(i)] of the patterned or substituted-mass Cantor set [Fig.~\ref{fig:hypercubic_illustration}(a,i)], diagnosed by the polarization $p_x$ in the $x$ direction [Eq.~\eqref{eq:polarization1}] over the $(M_0, M_{\rm alt})$ plane, within the range of $-1 \leq M_0 \leq 5$ and $-1 \leq M_{\rm alt} \leq 5$. The blue region is topological with $p_x=0.5$ and the white region is trivial with $p_x=0$. Each column corresponds to a specific generation ($g$) of the Cantor set, with $g=2$ [(a), (d), and (g)], $g=3$ [(b), (e), and (h)], and $g=4$ [(c), (f), and (i)]. Each row corresponds to a fixed total length ($L$) of the underlying 1D chain, with $L=81$ [(a)-(c)], $L=243$ [(d)-(f)], and $L=405$ [(g)-(i)]. At the second generation, the topological region nearly fills the parameter square except for trivial corners, while with increasing generation it contracts toward a horizontal strip, consistent with the growing relative weight of the red site or $M_{\rm alt}$ sector. At a fixed generation the phase boundaries are jagged for small $L$, but smooth into a stable shape as $L$ increases. (j) The topological fraction ($p$) of the phase diagram, defined as the fraction of sampled parameter points with $p_x=0.5$, as a function of $1/L$ for each generation, with dashed linear fits extrapolated to $1/L\to 0$, yielding its saturation value of $p_{\rm sat}$ in the thermodynamic limit. The offsets ($p_{\rm sat}$) between different generations indicate distinct large-$L$ limiting fractions of the topological regime, suggesting that the Su-Schrieffer-Heeger topological insulator can be realized in the thermodynamic limit ($1/L\rightarrow0$) in any generation. 
    The joint local density of states [(k)-(l)] associated with two particle-hole symmetric closest to zero-energy modes at energies $E_\pm$ (see legend) for the substituted-mass-fractal patterns on a $g=4$ Cantor set with $L=405$ with (I) open and (II) periodic boundary conditions. Chosen mass parameter values and the polarization of the resulting phase are: (k) $(M_0,M_{\rm alt},p_x)=(0.50,3.50,0.50)$ (l) $(M_0,M_{\rm alt},p_x)=(0.93,-0.05,0.50)$, (m) $(M_0,M_{\rm alt},p_x)=(2.00,-0.50,0.00)$, (n) $(M_0,M_{\rm alt},p_x)=(-0.50,2.00,0.50)$, and (o) $(M_0,M_{\rm alt},p_x)=(-0.75,0.05,0.00)$. The location of these points in the phase diagram from panel (i) are marked with green shapes. Grey vertical stripes show the sites with mass $M_0$ while white vertical stripes show sites with mass $M_{\rm alt}$. See Table~\ref{tab:cantor_values} for a summary on such choices of $M_0$ and $M_{\rm alt}$ values, and Sec.~\ref{subsec:1D} for discussion. 
    }~\label{fig:cantor_phase_diagram}
\end{figure*}

Starting from a 1D chain (with all sites colored black), the first-generation ($g=1$) Cantor set is constructed by identifying the central one-third of the lattice sites (colored red), which are either removed in method (a), integrated out in method (b), or assigned an on-site mass $M_{\rm alt}$ in method (c), while the remaining black sites retain an on-site mass $M_0$. To construct the second-generation ($g=2$) Cantor set, the same procedure is repeated on the two remaining segments of black sites. Iterating this procedure generates Cantor sets of arbitrary generation $g$ [Fig.~\ref{fig:hypercubic_illustration}(a)]. Since each generation is obtained by dividing every retained segment into three equal parts and removing the central third, the resulting Cantor set has a Hausdorff dimension $d_{\rm H}=\ln2/\ln3=\log_{3}2 \approx 0.631<1$. We now turn to the corresponding results.

Since the Cantor set is embedded in one dimension, methods (a) and (b) yield disconnected segments of 1D chains at every generation. Consequently, a global topological invariant, namely the polarization, cannot be computed reliably in these cases, as its evaluation requires PBCs. Notably, owing to the unique geometry of the Cantor set, even when PBCs are imposed on the parent 1D chain, its interior segments acquire OBCs at every generation. We therefore rely instead on a local topological marker, which is well defined for class BDI in $d=1$ and can be computed as follows.

Upon diagonalizing $H_{\rm SE}$ [Eq.~\eqref{eq:siteeliminated}] or $H_{\rm R}$ [Eq.~\eqref{eq:renormalized}], we construct the projector ${\mathbf P}$ onto the half-filled system, with all negative-energy states occupied:
\begin{equation}~\label{eq:projector}
{\mathbf P}= \sum_{E_i<0} \ket{\Psi_i} \bra{\Psi_i},
\end{equation}
where $\ket{\Psi_i}$ denotes an eigenstate of $H_{\rm SE}$ or $H_{\rm R}$ with energy $E_i$. Its complement, ${\mathbf Q}={\mathbf I}-{\mathbf P}$, where ${\mathbf I}$ is the identity matrix, projects onto the unoccupied positive-energy states. The local topological marker at position $x_i$ of the $i$th site is then given by the corresponding ($i$th) diagonal element of the operator~\cite{universalmarker}
\begin{equation}~\label{eq:utm_1d}
    {\mathbf C}_{\rm 1D}=N_D \; W \; \left[ {\mathbf Q} \; {\mathbf X} \; {\mathbf P} + {\mathbf P} \; {\mathbf X} \; {\mathbf Q} \right],
\end{equation}
which we evaluate under OBCs imposed on the parent 1D chain, inherited by the Cantor set. Here, $N_D=1$ is a normalization constant, $W=\tau_3$ is the generator of the sublattice symmetry, and ${\mathbf X}$ is the (diagonal) position operator with matrix elements ${\rm X}_{ij}=x_i \delta_{ij}$, where $\delta_{ij}$ is the Kronecker delta symbol. Within the topological regime of the SSH model, the local topological marker takes a quantized value of $1.00$ (within numerical errorbars) on the interior sites of each Cantor-set segment. By contrast, it vanishes on all sites for $M>4$ or $M<0$, corresponding to the topologically trivial regimes of the SSH model. These features are observed for both the site-elimination method [Fig.~\ref{fig:cantor_marker}(a)] and the projection method [Fig.~\ref{fig:cantor_marker}(d)]. Together, these findings establish the realization of SSH topology on the Cantor set, which we further corroborate through the bulk-boundary correspondence.

With OBCs and within the topological parameter regime of the SSH model, we find near-zero-energy modes whose corresponding local density of states (LDOS) is peaked at the two ends of each isolated segment of the Cantor set in method (a). By contrast, in the trivial regime, the LDOS of such modes is strongly localized near the center of each segment. See Fig.~\ref{fig:cantor_marker}(b). When PBCs are imposed on the parent 1D chain, the LDOS of the two modes closest to zero energy in the topological regime becomes peaked at the endpoints of the last Cantor-set segment, whose length is now twice that of the other segments owing to the PBC. In the trivial regime, however, the LDOS of these modes remains nearly uniform throughout the Cantor set. See Fig.~\ref{fig:cantor_marker}(c). Taken together, these observations demonstrate the compatibility of SSH topology on the Cantor set with the bulk-boundary correspondence in method (a).

In method (b), the second term of $H_{\rm R}$ [Eq.~\eqref{eq:renormalized}] generates long-range couplings among sites belonging to different segments of the Cantor set, originating entirely from $H^{-1}_{{\color{red}\bullet\bullet}}$, which is a non-sparse operator, even though $H_{{\color{red}\bullet\bullet}}$ itself is sparse, containing only on-site and nearest-neighbor hopping terms. Consequently, under OBCs and in the topological regime, the LDOS of the two modes closest to zero energy is strongly peaked at the two outermost endpoints of the Cantor set, as the long-range couplings connect the interior endpoints. In the trivial regime, by contrast, the LDOS of these modes is nearly uniform throughout the fractal lattice [Fig.~\ref{fig:cantor_marker}(e)]. Upon imposing PBCs, the LDOS of such modes in the topological regime becomes strongly peaked at the center of the longest segment of the Cantor set. In the trivial regime, however, the LDOS for such modes remains nearly uniform throughout the system [Fig.~\ref{fig:cantor_marker}(f)]. Taken together, these observations provide compelling evidence for the bulk-boundary correspondence of SSH topology on Cantor-set fractal lattices.

\begin{table}[]
    \centering
    \begin{tabular}{|c|c|c|c|c|}
    \hline
    $M_0$ value & $M_{\rm alt}$ value & Overall Phase & Figure & Symbol \\ \hline \hline
    Topological & Topological & Topological & Fig.~\ref{fig:cantor_phase_diagram}(k) & \markercircle      \\ \hline
    Topological & Topological & Normal & Fig.~\ref{fig:cantor_phase_diagram}(l) & \markersquare      \\ \hline
    Topological & Trivial & Topological & Fig.~\ref{fig:cantor_phase_diagram}(m) & \markertriangle    \\ \hline
    Topological & Trivial & Normal & Fig.~\ref{fig:cantor_phase_diagram}(n) & \markerinvtriangle \\ \hline
    Trivial & Topological & Topological & Fig.~\ref{fig:cantor_phase_diagram}(o) & \markerstar        \\ \hline
    Trivial & Topological & Normal & ---  & ---                \\ \hline
    Trivial & Normal & Topological & ---  & ---                \\ \hline
    Trivial & Normal & Normal & ---  & ---                \\ \hline
    \end{tabular}
    \caption{Selected $(M_0,M_{\rm alt})$ phase space points in Fig.~\ref{fig:cantor_phase_diagram} for the 1D patterned Cantor set [Fig.~\ref{fig:hypercubic_illustration}(a,i)]. The first and second columns show the values of the mass parameters $M_0$ (on black sites) and $M_{\rm alt}$ (on red sites), drawn from the topological or trivial parameter regimes of the Su–Schrieffer–Heeger model [Eq.~\eqref{eq:ham_1d}], respectively. The third column shows whether the resulting insulating phase is topological or normal, on a fourth generation patterned or substituted-mass Cantor set fractal lattice, determined from the polarization value in the $x$ direction, as also indicated in Fig.~\ref{fig:cantor_phase_diagram}(i) by the location of the respective symbol (fifth column) in the phase diagram. The fourth column refers to the corresponding subfigure of Fig.~\ref{fig:cantor_phase_diagram}, (k)-(o), displaying the local density of states associated to the two closest to zero-energy modes with open and periodic boundary conditions. Combinations with `---' in the fourth and fifth columns are either not found (third last row) or not possible (second last row) or exhibit no interesting bulk-boundary correspondence (last row).  
    }
    \label{tab:cantor_values}
\end{table}

Finally, we focus on method (c) to establish SSH topology on patterned Cantor set fractal lattices. As in this construction, no site gets eliminated or integrated out, one can impose PBCs on the underlying substituted-mass 1D chain and safely compute the polarization in the following way. The polarization in the $x$ direction is defined as~\cite{polariaztion:1, polarization:2} 
\allowdisplaybreaks[4]
\begin{equation}~\label{eq:polarization1}
p_x = n - n_{\rm al} \quad (\text{modulo} \:\: 1),
\end{equation}
where 
\allowdisplaybreaks[4]
\begin{equation}
n = -\operatorname{Re} \left[\frac{i}{2\pi}\, \operatorname{Tr}\!\left( \ln\!
\left\{ U^\dagger \exp\!\left[ \frac{2\pi i}{L}\sum_x x\,\hat{n}(x) \right] U \right\} \right) \right]
\end{equation}
$L$ is the length of the 1D chain, $\hat{n}(x)$ is the number operator at site $x$, $n_{\rm al} = \left(\sum_x x \right)/(2L)$, and at half filling $U$ is constructed by columnwise arranging all the eigenvectors of the filled states at negative energies. The topological (trivial) phase corresponds to $p_x=0.5$ ($p_x=0$).

Next, we construct the phase diagram of the patterned Cantor set as a function of $p_x$ over the $(M_0,M_{\rm alt})$ plane, revealing clear signatures of both topological and trivial insulating phases. We further observe that, at a fixed generation, the phase boundaries separating these phases become progressively smoother with increasing $L$. Moreover, the topological fraction of the phase diagram ($p$), when compared against $1/L$, approaches a finite and stable value $p_{\rm sat}$ upon extrapolation to the thermodynamic limit ($1/L \to 0$). While the second- and third-generation patterned Cantor sets exhibit broad two-parameter regions of topological insulating phases, the fourth-generation pattern supports topology within a horizontal striplike region. Thus, the substituted-mass construction provides a controlled setting in which generation-dependence and finite-size effects can be examined separately. We find that, $p_{\rm sat}$ is finite for every generation, but it decreases monotonically with increasing $g$. We therefore conclude that topological phases can be realized in patterned Cantor sets of arbitrary generation in the thermodynamic limit, although they become progressively less prevalent as $g$ increases. These results are summarized in Figs.~\ref{fig:cantor_phase_diagram}(a)-\ref{fig:cantor_phase_diagram}(j).

To relate these observations to the bulk-boundary correspondence, we consider eight possible scenarios arising from the two independent choices of $M_0$ and $M_{\rm alt}$, each of which can lie in either the topological or trivial parameter regime of the SSH model, together with the two possible outcomes for the resulting phase being topological or trivial. All eight possibilities are listed in Table~\ref{tab:cantor_values}, although only five are physically realizable. For each of these five cases, we examine the LDOS of the two modes closest to zero energy under both OBCs and PBCs. The results are presented in Figs.~\ref{fig:cantor_phase_diagram}(k)-\ref{fig:cantor_phase_diagram}(o).

When both $M_0$ and $M_{\rm alt}$ lie in the topological regime and the resulting phase is also topological, the LDOS of such modes is peaked at the outer endpoints (interior locations) of the system under OBCs (PBCs). The LDOS profiles under OBCs remain qualitatively unchanged even when the resulting phase is trivial, possibly highlighting the importance of the local topological environment, as the regions characterized by a topological $M_0$ terminate at interfaces with the vacuum. When $M_0$ and $M_{\rm alt}$ lies in the topological and trivial regime, respectively, the LDOS under OBCs is peaked at the endpoints (in the interior) when the resulting phase is topological (trivial). Under PBCs, however, it is consistently peaked near an internal interface between the black and red sites, further emphasizing the role of the local topological environment. By the same token, finally, when $M_0$ ($M_{\rm alt}$) lies in the trivial (topological) regime and the resulting phase is topological, the LDOS of such a pair of modes is consistently peaked at the interfaces between black and red sites. The remaining three combinations are either not observed, as in the case where $M_0$ ($M_{\rm alt}$) lies in the trivial (topological) regime and the resulting phase is trivial, or are physically forbidden (uninteresting), as in the case where both $M_0$ and $M_{\rm alt}$ lie in the trivial regime but the resulting phase is topological (trivial).

\begin{figure}[t!]
    \centering
    \includegraphics[width=0.925\linewidth]{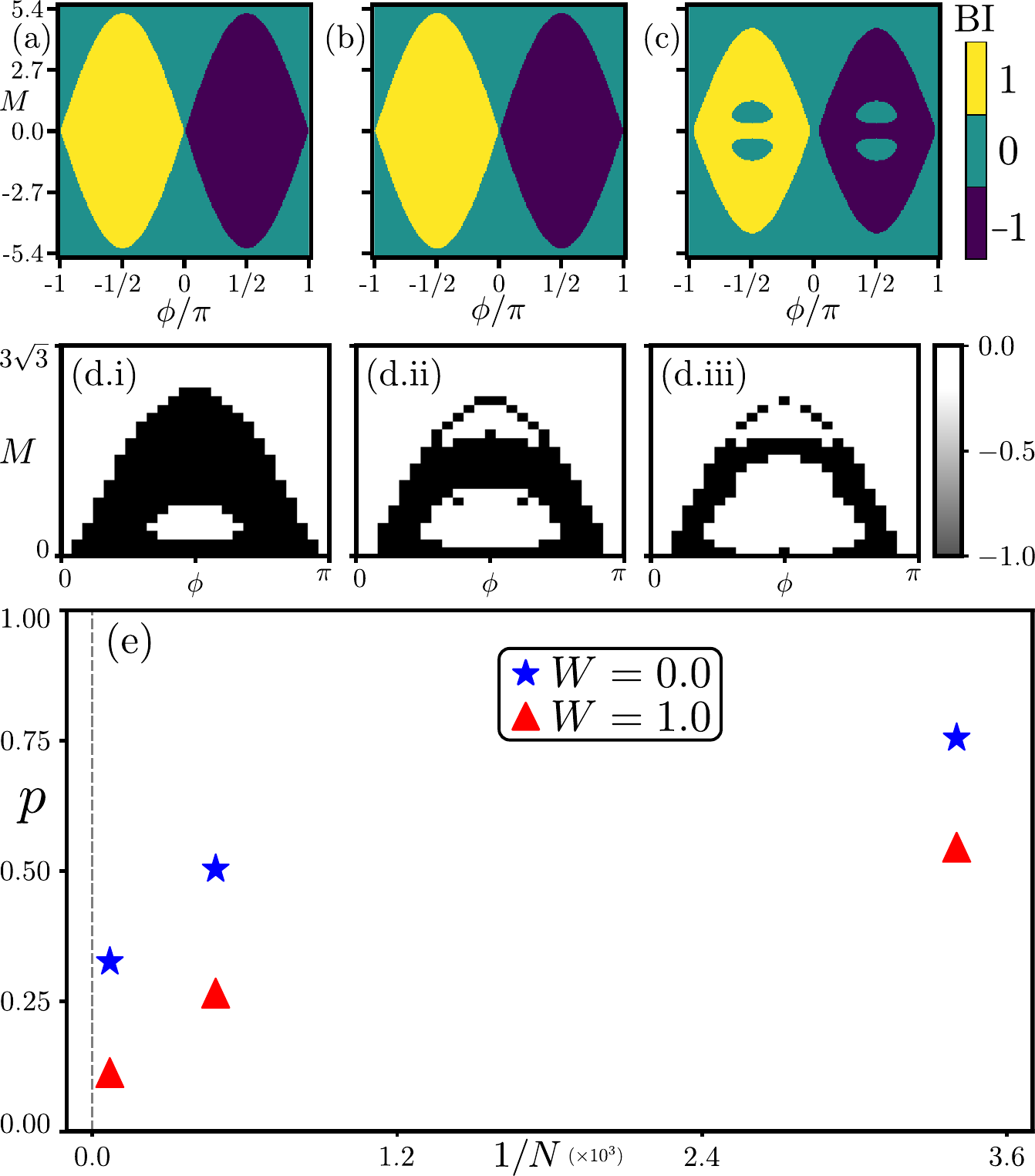}
    \caption{Phase diagram of the Haldane model [Eq.~\eqref{eq:Haldane}] in term of the quantized Bott index [Eq.~\eqref{eq:BottIndex}] of ${\rm BI}=0, \pm 1$ over the $(\phi, M)$ parameter space, with $-\pi \leq \phi \leq \pi$ and
    $-5.4 \leq M \leq 5.4$ on (a) a parent honeycomb lattice, containing 3282 sites and on its geometric descendant second generation hexaflake fractals, containing 2058 sites, obtained via (b) method of projection or renormalization and (c) method of site elimination, for $t_1=1$. On the parent honeycomb lattice the phase boundaries between the topological and normal insulators are determined by $M=\pm 3\sqrt{3}\, t_2\sin\phi$, with the latter one residing outside the topological lobes, with nearly identical outcomes in (b). In (c) the topological lobes remain recognizable, but are reduced in area relative to (a) and (b), namely the lobes are now shortened, two lobes no longer meet at $\phi=0$, the
    regions near $\phi=\pm\pi$ at $M=0$ become trivial, and internal trivial holes open within each topological lobe. (d) Phase diagram of the Haldane model on hexaflake lattice of generation (i) two, (ii) three, and (iii) four within a restricted parameter regime of $0 \leq \phi \leq \pi$ and $0 \leq M \leq 5.4$ in the site elimination method. (e) Fraction of the topological regime ($p$) within the parameter regime from (d) as a function of $1/N$, where $N$ is the number of lattice sites in the hexaflake in clean (red) system ($W=0$) and in the presence of disorder (blue) with $W=1.00$, discussed in Sec.~\ref{sec:disorder}, suggesting a finite $p$, namely $p_{\rm sat}$ as $1/N \to 0$ (thermodynamic limit). 
    }
    \label{fig:haldane_phase_diagram}
\end{figure}

\subsection{Two dimensions}~\label{subsec:2D}

We begin the discussion of possible realizations of topological insulators on 2D fractal lattices by considering the Haldane model on the parent honeycomb (or hexagonal) lattice and subsequently demonstrating its manifestations on the geometric descendant hexaflake network [Fig.~\ref{fig:haldane_setup_figure}(b)]. The Hausdorff dimension of the hexaflake lattice is obtained from its self-similar construction, in which each iteration generates seven smaller copies of the parent hexagon with each arm being scaled by a factor of 1/3. Accordingly, the Hausdorff dimension is given by $d_{\rm H}=\ln 7/\ln 3 = \log_7 3 \approx 1.771<2$. The tight-binding Hamiltonian of the Haldane model reads as~\cite{haldane1988}
\begin{equation}~\label{eq:Haldane}
H^{\rm TB}_{\rm Hal}
= t_1 \sum_{\langle i,j\rangle} c_i^\dagger c_j
+ t_2 \sum_{\langle\!\langle i,j\rangle\!\rangle} e^{i\nu_{ij}\phi}\, c_i^\dagger c_j
+ M \sum_i \xi_i\, c_i^\dagger c_i,
\end{equation}
where $c^\dagger_i$ ($c_i$) denotes the fermionic creation (annihilation) operator at the $i$th site. In this setup, the real nearest-neighbor hopping $t_1$, connecting sites belonging to complementary sublattices and denoted by $\langle \cdots \rangle$, gives rise to massless Dirac fermions at the six corners of the hexagonal Brillouin zone. The complex next-nearest-neighbor hopping $t_2 e^{i\nu_{ij}\phi}$ connects sites belonging to the same sublattice and is denoted by $\langle\!\langle \cdots \rangle\!\rangle$. Here, the sign $\nu_{ij}=\pm1$ is determined by the orientation of the hopping path, giving rise to a Haldane flux phase $\phi$ enclosed by each closed path. Although the net flux through each honeycomb plaquette vanishes, the system can support topological Chern or QAHI in which the nodal Dirac fermions acquire a mass. The last term in Eq.~\eqref{eq:Haldane}, with $\xi_i=\pm 1$ labeling the two triangular sublattices of the honeycomb lattice, generates a staggered fermion density between the two sublattices and represents the Semenoff mass~\cite{semenoff1984}. Figure~\ref{fig:haldane_setup_figure} shows pictorial representations of all the terms from Eq.~\eqref{eq:Haldane}.

To appreciate the symmetry class of $H^{\rm TB}_{\rm Hal}$, we define an $N$-component spinor $\Psi^\top=({\mathbf c}_A, {\mathbf c}_B)$, where $N$ is the the number of lattice sites in the system, ${\mathbf c}_A$ (${\mathbf c}_B$) is an $N/2$-component spinor constructed by the fermionic annihilation operators on the sites of $A$ ($B$) sublattice of the honeycomb lattice, and $\top$ stands for transposition. In this basis, the Hamiltonian operator associated with $H^{\rm TB}_{\rm Hal}$ takes the form 
\begin{equation}
\hat{h}^{\rm TB}_{\rm Hal} = \left( \begin{array}{cc}
{\mathbf t}_2+ {\mathbf M} & {\mathbf t}_1 \\
{\mathbf t}^\top_1 & -{\mathbf t}_2- {\mathbf M}
\end{array}
\right),
\end{equation}
where ${\mathbf t}_1$ (${\mathbf t}_2$) is an $N/2$-dimensional inter-sublattice (intra-sublattice) real (complex Hermitian) hopping matrix, proportional to $t_1$ ($t_2$), and ${\mathbf M}$ is proportional to an $N/2$-dimensional identity matrix, with magnitude $M$. For $t_2=M=0$, the system describes a collection of massless Dirac fermions and $\hat{h}^{\rm TB}_{\rm Hal}$ preserves the time-reversal symmetry, generated by ${\mathcal T}={\mathcal K}$, particle-hole symmetry, generated by ${\mathcal C}=\Sigma_z {\mathcal K}$, and sublattice symmetry under ${\mathcal S}=\Sigma_z$, where $\Sigma_z={\rm diag}. \left( {\mathbf I}_{N/2}, - {\mathbf I}_{N/2} \right)$ with ${\mathbf I}_{N/2}$ as an $N/2$-dimensional identity matrix, thus belonging to class BDI. All three symmetries are broken for finite $t_2$ and $M$, and the system enters into class A.

Since the Semenoff mass trivially gaps out the Dirac fermions, it competes with the Haldane mass, giving rise to a confluence of topological and trivial insulators in the $(\phi, M)$ parameter space. While the topological insulating phase is characterized by a first Chern number $C=\pm 1$, its trivial counterpart has $C=0$~\cite{TKNN}. The phase boundaries separating these two phases are given by
\begin{equation}\label{eq:haldanebndy}
M = \pm 3\sqrt{3}\, t_2 \sin\phi.
\end{equation}
Throughout, we set $t_1=t_2=1$. Consequently, the topological insulating phase survives up to a maximum value of $|M|=3\sqrt{3}\approx 5.196$, realized for $\phi=\pi/2$. We therefore scan the phase diagram over the parameter regime $|M|\leq 5.4$ and $|\phi| \leq \pi$. We construct the phase diagrams of the Haldane model on the parent honeycomb lattice and its geometrically derived hexaflake descendants by computing the Bott index. This is achieved by numerically diagonalizing $H^{\rm TB}_{\rm Hal}$ and either $H_{\rm SE}$ [method (a)] or $H_{\rm R}$ [method (b)], respectively. Note that $H_{\rm SE}$ or $H_{\rm R}$ can be readily obtained by casting $H^{\rm TB}_{\rm Hal}$ into the block-matrix form of $H_{\rm parent}$, given in Eq.~\eqref{eq:blockmatrix}, and subsequently following the procedure described in Sec.~\ref{subsec:methods}.

\begin{figure}[t!]
    \centering
    \includegraphics[width=1.00\linewidth]{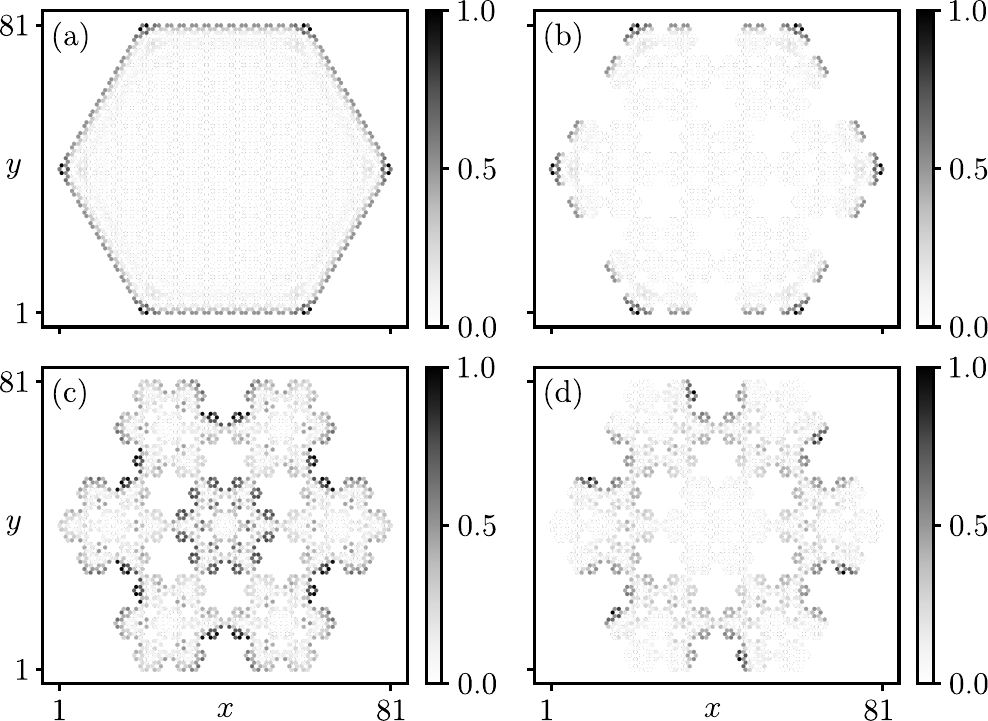}
    \caption{Joint local density of states (normalized by its maximal value) of the two particle-hole symmetric closest-to-zero energy modes for the Haldane model, defined on a parent honeycomb lattice with 3282 sites, $\phi=\pi/2$, and $M=3\sqrt{3}/2$ [Eq.~\eqref{eq:Haldane}] on (a) the honeycomb lattice, on its geometric descendant third generation hexaflake fractal with (b) method of projection or renormalization and (c) method of site elimination with open boundary conditions. Panel (d) is same as (c) but with periodic boundary conditions. The energy eigenvalues of such modes are $\pm E$ with (a) $E =1.85\times10^{-2}$, (b) $E=2.96\times10^{-2}$, (c) $E =2.64\times10^{-2}$, and (d) $E =2.10\times10^{-2}$.
    }~\label{fig:haldane_ldos}
\end{figure}

In the absence of the translational symmetry, as is the case for the fractal lattices, the notion of Brillouin zone and the first Chern number becomes moot. We must therefore turn to an alternative topological invariant, which in this case is the Bott Index. To compute the Bott index, we define two diagonal matrices ${\mathbf X}$ and ${\mathbf Y}$ whose elements are ${\rm X}_{ij}=x_i \delta_{ij}$ and ${\rm Y}_{ij}=y_i\delta_{ij}$, respectively, encoding the position of the $i$th site in the $x$ and $y$ directions. Subsequently, we construct two unitary operators ${\mathbf U}_{\rm X}=\exp\left( 2\pi i {\mathbf X}\right)$ and ${\mathbf U}_{\rm Y}=\exp\left( 2\pi i {\mathbf Y}\right)$. The topological invariant of the Haldane model, namely the Bott index, is defined as~\cite{HastingLoring}
\begin{equation}~\label{eq:BottIndex}
{\rm BI}=\frac{1}{2 \pi} \Im \Big({\rm Tr}\Big[\ln ({\mathbf V}_{\rm X} \; {\mathbf V}_{\rm Y} \; {\mathbf V}_{\rm X}^\dagger \; {\mathbf V}_{\rm Y}^\dagger)\Big]\Big),
\end{equation}
where ${\mathbf V}_{\rm A}={\mathbf Q} + {\mathbf P} {\mathbf U}_{\rm A} {\mathbf P}$ for ${\rm A}={\rm X}$ and ${\rm Y}$ with ${\mathbf P}$ as the projector in the half filled system with all the states at negative energies being filled, defined in Eq.~\eqref{eq:projector}, and ${\mathbf Q}$ as its complement. It should, however, be noted that while computing the Bott index for the honeycomb lattice [hexaflake fractal lattices], $\ket{\Psi_i}$s are eigenstates with negative eigenvalues $E_i$s of $H^{\rm TB}_{\rm Hal}$ [$H_{\rm SE}$ in method (a) or $H_{\rm R}$ in method (b)]. Next we discuss the phase diagrams associated with the Haldane model on the parent honeycomb lattice and hexaflake fractal lattices over the $(\phi, M)$ space. Results are shown in Fig.~\ref{fig:haldane_phase_diagram}. The Bott index is always computed by imposing PBCs on $H_{\rm parent}$ in all directions, which inherited by $H_{\rm SE}$ and $H_{\rm R}$.

\begin{figure*}[t]
    \centering
    \includegraphics[width=0.95\linewidth]{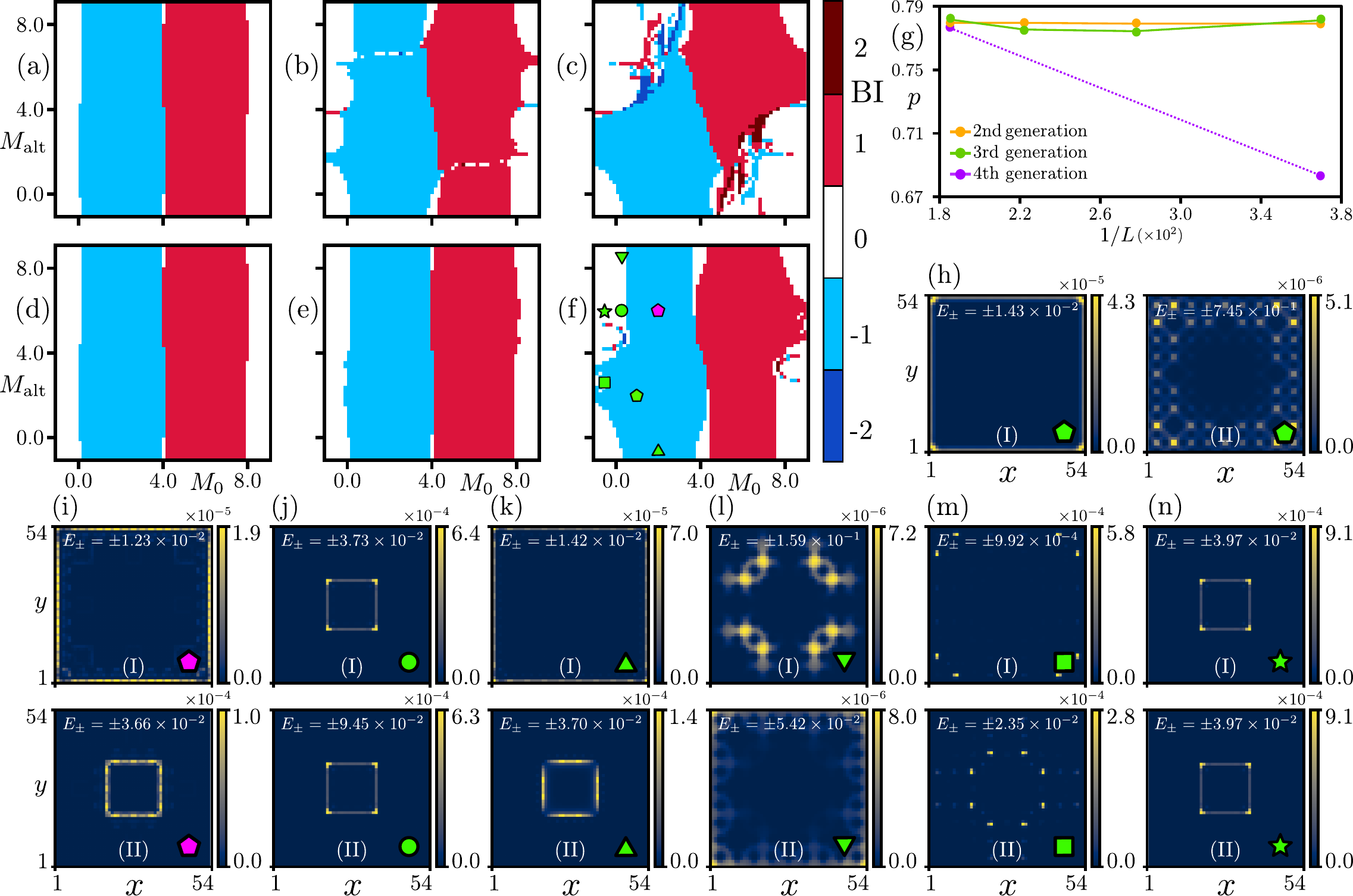}
    \caption{Topology of the patterned or substituted-mass Sierpi\'nski carpet fractal lattices, diagnosed by the
    Bott index (${\rm BI}$). Panels (a)--(f) show phase diagrams over the $(M_0, M_{\rm alt})$ plane in terms of the Bott index with the color bar spanning ${\rm BI}=-2,-1,0,1,2$. The columns correspond to the generation number of $g=1$ [(a) and (d)], $g=2$ [(b) and (e)], and $g=3$ [(c) and (f)] of the patterned carpet, and the rows correspond to linear size of $L=27$ [(a)-(c)] and $L=54$ [(d)-(f)] of the parent square lattice. Panels (a), (d), and (e) closely resemble the conventional $M_{\rm alt}=M_0$ case, in which the two topological regions $0<M_0<4$ and $4<M_0<8$ appear as strips with a
    weak dependence on $M_{\rm alt}$; while the boundary between them at $M_0=4$ gets distorted in panels (c) and (f), showing increased dependence on $M_{\rm alt}$, and isolated patches carrying $B=\pm2$ phases. Panel (c) ($g=3$ and $L=27$) is the most strongly modified, with distorted boundaries and small extended $B=\pm2$ regions, which, however, progressively disappear from the phase diagram with increasing $L$. (g) The topological fraction $p$, defined as the fraction of sampled parameter points with ${\rm BI} \neq 0$ (counting all of ${\rm BI}=\pm1,\pm2$), as a function of $1/L$ for each generation. The first- and second-generation curves contain four data points each, while the third generation has only two owing to the computational cost of larger systems. The first-generation curve sits just below $p=0.8$, the value expected for a purely vertical topological region spanning $0<M_0<8$ within the plotted mass range. The second generation curve shows almost no dependence on $L$. The third generation curve shows an upturn from $p \approx 0.68$ to $p \approx 0.78$ as $L$ is increased from $27$ to $54$, suggesting stability of topological phases of substituted-mass Sierpi\'nski carpet fractal lattices in any generation in the thermodynamic limit ($1/L \to 0$). Columns (h)--(m) show the local density of states for two particle-hole symmetric closest-to-zero-energy modes at energies $E_\pm$ (see legend) on the $g=3$ and $L=54$ substituted-mass Sierpi\'nski carpet fractal lattices with (I) open and (II) periodic boundary conditions. Chosen mass parameter values and the BI of the resulting phases are: (h) $(M_0, M_{\rm alt},{\rm BI})=(1.00, 2.00,1.00)$, (i) $(M_0,M_{\rm alt},{\rm BI})=(2.00, 6.00,1.00)$, (j) $(M_0,M_{\rm alt},{\rm BI})=(0.25, 6.00,0.00)$, (k) $(M_0,M_{\rm alt},{\rm BI})=(2.00, -0.50,1.00)$, (l) $(M_0,M_{\rm alt},{\rm BI})=(0.25, 8.50,0.00)$, (m) $(M_0,M_{\rm alt},{\rm BI})=(-0.50, 2.50,1.00)$, and (n) $(M_0,M_{\rm alt},{\rm BI})=(-0.50,6.00,0.00)$. The location of these points in the phase diagram in panel (f) are marked with green and magenta shapes. Reasoning behind the choice of such parameter values is summarized in Table~\ref{tab:carpet_values} and discussed in Sec.~\ref{subsec:2D}.}
    \label{fig:carpet_phase_diagram}
\end{figure*}

On the honeycomb lattice, the computed Bott index yields a phase diagram identical to that previously reported using the first Chern number, obtained by integrating the Berry curvature over the first Brillouin zone [Fig.~\ref{fig:haldane_phase_diagram}(a)]. Specifically, the phase diagram in the $(\phi, M)$ plane exhibits two lobes of topological phases with quantized Bott indices of opposite signs (${\rm BI}=\pm 1$), surrounded by trivial insulating phases with zero Bott index. The phase boundaries closely follow those given by Eq.~\eqref{eq:haldanebndy}, within numerical error bars. In contrast, the phase diagram for the hexaflake lattice obtained using the site elimination method [method (a)] exhibits an erosion of the topological regions. Specifically, (1) the topological lobes shrink in area on the $(\phi, M)$ plane, (2) holes corresponding to trivial insulating phases appear within each lobe, and (3) regions near the $M=0$ line become trivial in the vicinity of $\phi=0$ and $\pm\pi$. These features are shown in Fig.~\ref{fig:haldane_phase_diagram}(b). However, when using the projection or renormalization approach [method (b)], the resulting phase diagram is \emph{identical} to that of the honeycomb lattice, within numerical error bars, as shown in Fig.~\ref{fig:haldane_phase_diagram}(c). This result demonstrates the exactness of the Schur complement in preserving the topological properties of the parent system on its geometric descendant. In contrast, site elimination disrupts the stability of topological phases of matter.

We, therefore, study the phase diagram of the Haldane model in the site elimination method on the hexaflake lattice of varying generation number ($g$). The results are shown in Fig.~\ref{fig:haldane_phase_diagram}(d) for $g=2,3$, and $4$. With increasing $g$ the trivial hole within the topological lobe grows in area. To further scrutinize the stability of the topological regime within the site elimination framework, we track how the topological fraction ($p$) of the phase space in the $(\phi, M)$ plane evolves with increasing fractal generation ($g$). As shown in Fig.~\ref{fig:haldane_phase_diagram}(e), $p$ decreases with increasing $g$. This trend is physically expected, as repeated site elimination renders the honeycomb-derived hexaflake structure increasingly sparse. Consequently, a Chern insulator, strictly defined in $d=2$, is expected to become less stable on such a self-similar network. However, this dataset must be interpreted with the caveat that the generation number and system size are varied simultaneously. Thus, Fig.~\ref{fig:haldane_phase_diagram}(e) does not disentangle the effects of increasing fractality from ordinary finite-size effects. Nevertheless, even within this restricted scope, the extrapolated trend suggests a finite $p_{\rm sat}$, indicating that site elimination procedure leaves a finite region of parameter space supporting topological insulating phases with quantized Bott indices in the thermodynamic limit when the number of sites on the hexaflake fractal $N \to \infty$.

Finally, we establish a one-to-one correspondence between the topological insulating phases arising from the Haldane model and their associated bulk-boundary correspondence, manifested through near-zero-energy edge modes. On the honeycomb lattice, we observe edge-localized near-zero-energy modes with OBCs, as shown in Fig.~\ref{fig:haldane_ldos}(a). This observation remains qualitatively unchanged on the hexaflake fractal lattice when its effective Hamiltonian is constructed using the renormalization or projection method, as shown in Fig.~\ref{fig:haldane_ldos}(b). Notably, despite possessing inner edges, the hexaflake lattice exhibits no appreciable LDOS weight associated with the near-zero-energy modes along its inner edges. This is because the sites on such edges are connected to all other sites in the system through long-range hopping, generated by $H^{-1}_{{\color{red}\bullet\bullet}}$ (a non-sparse operator), which enters the corresponding Hamiltonian $H_{\rm R}$. By stark contrast, under OBCs, the LDOS weight of the near-zero-energy modes of $H_{\rm SE}$ (site elimination method) is distributed over both the outer and inner edges of the hexaflake fractal lattice, as shown in Fig.~\ref{fig:haldane_ldos}(c). Remarkably, when PBCs are imposed on $H_{\rm parent}$ during the construction of $H_{\rm SE}$, only the inner edges become exposed, harboring prominent LDOS peaks associated with the topological modes closest to zero energy, as shown in Fig.~\ref{fig:haldane_ldos}(d).

\begin{table}[t!]
    \centering
    \begin{tabular}{|c|c|c|c|c|}
    \hline
    $M_0$ value & $M_{\rm alt}$ value & Overall Phase & Figure & Symbol \\ \hline \hline
    Topo ($\Gamma$) & Topo ($\Gamma$) & Topological & Fig.~\ref{fig:carpet_phase_diagram}(h) & \markerpentagon  \\
    Topo ($\Gamma$) & Topo (${\rm M}$) & Topological & Fig.~\ref{fig:carpet_phase_diagram}(i) & \markerpentagon[markermagenta] \\ \hline
    Topo & Topo & Normal & Fig.~\ref{fig:carpet_phase_diagram}(j) & \markercircle      \\ \hline
    Topo & Trivial & Topological & Fig.~\ref{fig:carpet_phase_diagram}(k) & \markertriangle    \\ \hline
    Topo & Trivial & Normal & Fig.~\ref{fig:carpet_phase_diagram}(l) & \markerinvtriangle \\ \hline
    Trivial & Topo & Topological & Fig.~\ref{fig:carpet_phase_diagram}(m) & \markersquare      \\ \hline
    Trivial & Topo & Normal & Fig.~\ref{fig:carpet_phase_diagram}(n) & \markerstar        \\ \hline
    Trivial & Trivial & Topological & ---  & ---                \\ \hline
    Trivial & Trivial & Normal & ---  & ---                \\ \hline
    \end{tabular}
    \caption{Selected $(M_0,M_{\rm alt})$ phase space points in Fig.~\ref{fig:carpet_phase_diagram} for the 2D patterned or substituted-mass Sierpi\'nski carpet [Fig.~\ref{fig:hypercubic_illustration}(a,i)]. The first and second columns show the values of the mass parameters $M_0$ (on black sites) and $M_{\rm alt}$ (on red sites), drawn from the topological (Topo) or trivial parameter regimes of the Qi–Wu–Zhang model [Eq.~\eqref{eq:qzw}], respectively. The third column shows whether the resulting insulating phase is topological or normal, on a third generation patterned or substituted-mass Sierpi\'nski carpet fractal lattice embedded on a $L=54$ parent square lattice, determined from the Bott index (BI), as also indicated in Fig.~\ref{fig:carpet_phase_diagram}(f) by the location of the respective symbol (fifth column) in the phase diagram.  The fourth column refers to the corresponding subfigure of Fig.~\ref{fig:carpet_phase_diagram}, (h)-(n), displaying the local density of states associated to the two closest to zero-energy modes with open and periodic boundary conditions. Combinations with `---' in the fourth and fifth columns are either not possible (second last row) or exhibit no interesting bulk-boundary correspondence (last row). As the Chern insulating phase can emerge with a band inversion near the $\Gamma$ or ${\rm M}$ point, characterized by opposite Chern numbers or Bott indices, the overall phase can be topological with the possibility of drawing $M_0$ and $M_{\rm alt}$ from the parameter regimes associated with either of these two distinct phases (indicated by the symbol within the parentheses after `Topo').
    }
    \label{tab:carpet_values}
\end{table}

Next, we focus on the square-lattice-based QWZ model of QAHIs and demonstrate its characteristic topological signatures on patterned or substituted-mass Sierpi\'nski carpet fractal lattices. The Bloch Hamiltonian of the QWZ model on a square lattice with lattice spacing $a$ is given by~\cite{QWZ}
\begin{eqnarray}~\label{eq:qzw}
\hspace{-0.75cm}
H^{\rm QWZ}_{\rm 2D}(\bm{k})
&=& t\left[\sin(k_x a) \tau_1 + \sin(k_y a) \tau_2 \right] \nonumber \\
&+&\left[M-2B\left\{ 2-\cos(k_x a)-\cos(k_y a) \right\} \right] \tau_3,
\end{eqnarray}
where $\{ \tau_\mu \}$ denotes the set of Pauli matrices acting on the orbital degrees of freedom. The QWZ model realizes topological insulating phases for $0<M/B<8$ and a trivial insulating phase otherwise. Interestingly, despite its minimal two-band structure, the QWZ model supports two distinct topological phases owing to band inversions at different inversion-symmetric high-symmetry points of the Brillouin zone. Specifically, for $0<M/B<4$ and $4<M/B<8$, the band inversion occurs at the $\Gamma$ and ${\rm M}$ points, respectively, yielding QAHI phases characterized by a first Chern number of $\pm 1$~\cite{TKNN}.

The tight-binding Hamiltonian corresponding to the QWZ model [Eq.~\eqref{eq:qzw}] on the parent square lattice can be written as
\begin{eqnarray}~\label{eq:QWZTB}
H^{\rm QWZ}_{\rm 2D, TB} &=& \sum_{\vec{r}} \bigg\{ \bigg( \sum_{\mu=x,y} \Psi^\dagger_{\vec{r}} \left[ \frac{t}{2i} \tau_\mu + B \tau_3 \right] \Psi_{\vec{r}+\hat{e}_\mu} \nonumber \\
&+& {\rm H.c.} \bigg) 
+ \Psi^\dagger_{\vec{r}} \left( M+4 B \right) \tau_3 \Psi_{\vec{r}} \bigg\} + \cdots.
\end{eqnarray}
Here, $\Psi^\top_{\vec{r}}=\left( c_{\vec{r},+}, c_{\vec{r},-} \right)$ is a two-component fermionic spinor, where $c_{\vec{r},\tau}$ annihilates a fermion at site $\vec{r}=(x,y)$ in the orbital with parity eigenvalue $\tau=\pm$, and $\tau_{x,y}\equiv\tau_{1,2}$. The ellipsis denotes additional hopping terms required to implement PBCs, which are omitted for brevity. Throughout our numerical calculations, we set $t=B=1$. We subsequently construct the block-matrix Hamiltonian $H_{\rm parent}$ from $H^{\rm QWZ}_{\rm 2D, TB}$ by assigning distinct on-site mass parameters, $M_0$ and $M_{\rm alt}$, to the black and red sites, respectively, corresponding to the Sierpi\'nski carpet fractal lattice and its complement [see Fig.~\ref{fig:hypercubic_illustration}(b,i)]. We then investigate the topological phase diagram of the resulting patterned or substituted-mass Sierpi\'nski carpet in the $(M_0,M_{\rm alt})$ parameter space, characterizing the phases using the Bott index [Eq.~\eqref{eq:BottIndex}]~\cite{HastingLoring}.

The phase diagrams of the patterned Sierpi\'nski carpet for generations $g=1,2$, and $3$, and different linear system sizes $L$ of the parent square lattice are presented in Figs.~\ref{fig:carpet_phase_diagram}(a)-\ref{fig:carpet_phase_diagram}(f). In contrast to the previously discussed patterned Cantor set, the resulting phase diagrams exhibit comparatively modest deviations from the conventional QWZ model. Specifically, for $g=1$ and $2$, the topological regions remain nearly vertical, extending over the parameter range $0<M<8$ with only a weak dependence on $M_{\rm alt}$. A noticeable departure from this behavior emerges for the third-generation carpet and smaller system sizes, where the phase boundaries become curved and small islands with ${\rm BI}=\pm2$ appear. These islands are a finite-size effect associated with the substituted-mass Sierpi\'nski carpet, since the uniform QWZ model, corresponding to $M_0=M_{\rm alt}$, supports only ${\rm BI}=\pm1$ phases. Consistent with this interpretation, the ${\rm BI}=\pm2$ islands disappear upon increasing the system size within the range accessible to our numerical calculations.

Notably, the fraction of the topological region in the phase diagram ($p$) remains nearly independent of the system size for patterned Sierpi\'nski carpets of generations $g=1$ and $2$, suggesting a finite saturation value of $p$, namely $p_{\rm sat}$, in the thermodynamic limit ($1/L\to 0$). For $g=3$, only two data points are available in the $(p,1/L)$ plane, precluding a conclusive finite-size scaling analysis. Nevertheless, $p$ exhibits a clear increasing trend with system size, approaching values comparable to those obtained for $g=1$ and $2$ at the largest accessible $L$. This trend suggests that topological phases persist in patterned or substituted-mass Sierpi\'nski carpets across different generations in the thermodynamic limit. The corresponding results are presented in Fig.~\ref{fig:carpet_phase_diagram}(g).

\begin{figure*}[t!]
    \centering
    \includegraphics[width=\linewidth]{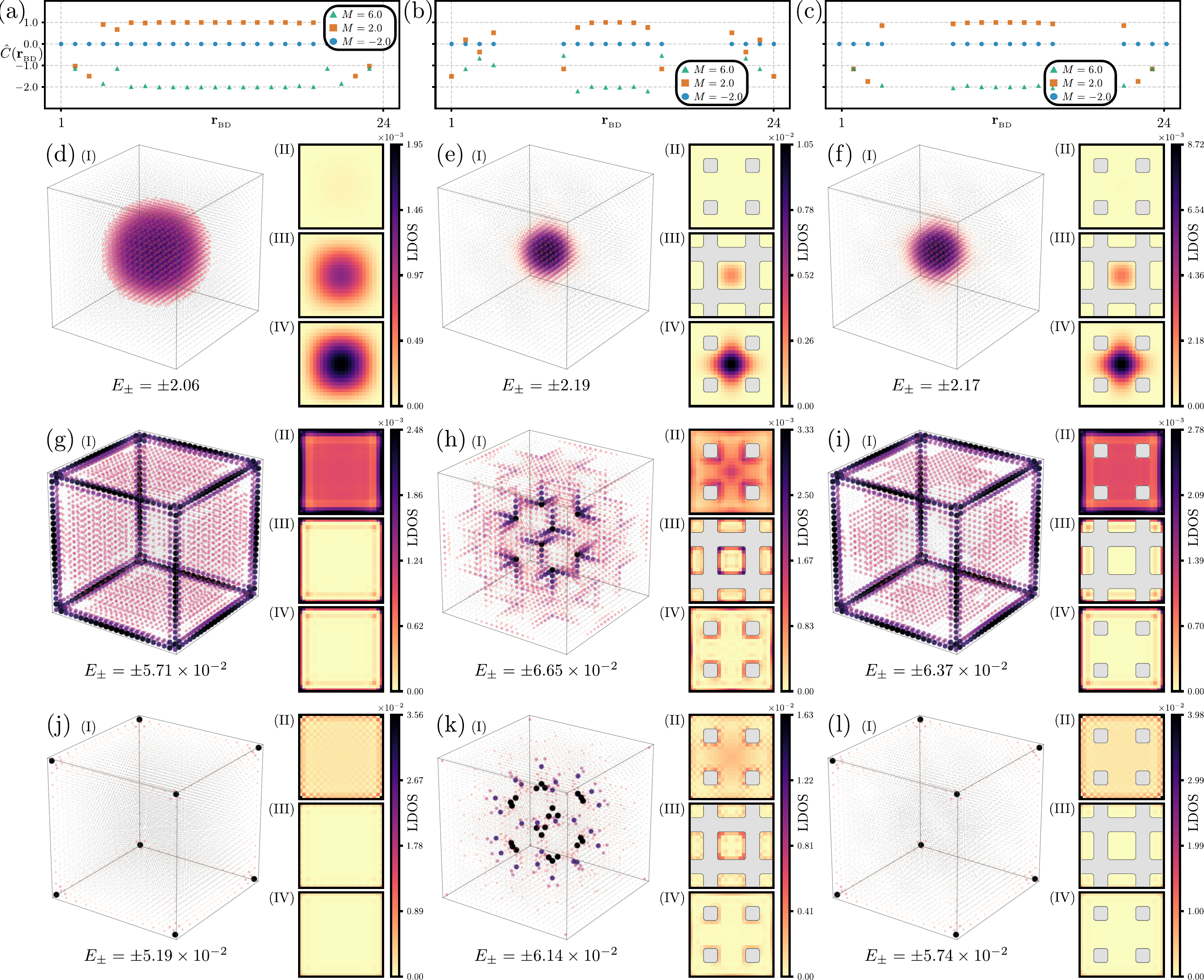}
    \caption{Local topological marker $C_{\rm 3D}(\vec{r})$ [Eq.~\eqref{eq:utm_3d}] on a $L=24$ (a) parent cubic lattice, and tiled Menger sponge obtained via the method of (b) site elimination [method 1], and (c) projection or renormalization [method 2] for mass values (see legends) of $M=-2.0$ (trivial insulator), and $M=2.0$ and $6.0$ (topological insulator), defined in Eq.~\eqref{eq:ham_3d}, shown on the sites of the body-diagonal ($\vec{r}_{_{\rm BD}}$). The local density of states (LDOS) associated with two particle-hole symmetric closest-to-zero-energy modes at energies $E_\pm$ (see legends) with open boundary conditions [(d)--(l)] on a $L=24$ parent cubic lattice [(d), (g), (j)], and tiled Menger sponge obtained via method of site elimination [(e), (h), (k)], and projection or renormalization [(f), (i), (l)] for $M=-2.0$ [(d)-(f)], $2.0$ [(g)-(i)], and $6.0$ [(j)-(l)] are shown in (I) where the largest 20\% LDOS values are projected onto the lattice while the remaining sites are masked to a grey color and on its two-dimensional slices at (II) $z=1$, (III) $z=7$, and (IV) $z=13$, where the holes are colored grey.
    }~\label{fig:sponge_marker}
\end{figure*}

We next establish the bulk-boundary correspondence in substituted-mass Sierpi\'nski carpets by examining the manifold of near-zero-energy modes under open and periodic boundary conditions. The corresponding results are presented in Figs.~\ref{fig:carpet_phase_diagram}(h)-\ref{fig:carpet_phase_diagram}(n), and also summarized in Table~\ref{tab:carpet_values}. The overall phenomenology closely resembles that previously observed for the substituted-mass Cantor set, with one notable distinction: the near-zero-energy modes are localized along the outer boundary and/or the largest internal boundaries surrounding the central regions of the fractal lattice, bordering the regions with distinct values of on-site masses $M_0$ and $M_{\rm alt}$. Furthermore, the possibility of distinct QAHI phases with opposite Bott indices gives rise to several interesting combinations of boundary-localized modes, which have no counterpart in the 1D SSH model. Next, we examine these possibilities.

For instance, when $M_0$ and $M_{\rm alt}$ are chosen from parameter regimes supporting the $\Gamma$ and ${\rm M}$ phases, respectively, the resulting system can realize either a topological or a normal insulating phase. In the former case, OBCs give rise to pronounced LDOS peaks along the outer edges, corresponding to a pair of particle-hole-symmetric near-zero-energy modes. Such outer-edge signatures are absent in the normal insulating phase, however. Interestingly, imposing PBCs along both directions reveals pronounced LDOS peaks associated with these modes near the central inner edges in both phases. These results are presented in Figs.~\ref{fig:carpet_phase_diagram}(i) and~\ref{fig:carpet_phase_diagram}(j), respectively. The behavior in Fig.~\ref{fig:carpet_phase_diagram}(i) is qualitatively similar to that observed when $M_0$ and $M_{\rm alt}$ are chosen from topological and trivial parameter regimes, respectively, such that the resulting phase is topological [Fig.~\ref{fig:carpet_phase_diagram}(k)]. In contrast, when $M_0$ and $M_{\rm alt}$ are chosen from trivial and topological parameter regimes, respectively, yielding a normal insulator, the near-zero-energy LDOS peaks are confined to the largest inner edges under both OBCs and PBCs [Fig.~\ref{fig:carpet_phase_diagram}(n)]. Taken together, these results establish a correspondence between the bulk topological invariant and the emergence of boundary-localized near-zero-energy modes, while highlighting the crucial role of the local topological environment in determining their spatial distribution.

\subsection{Three dimensions}~\label{subsec:3D}

Finally, we turn to three-dimensional (3D) topological insulators belonging to the symmetry class AII, realized on a cubic lattice, and investigate their topological signatures on a geometric descendant of the \emph{tiled} Menger sponge, constructed by arranging eight regular Menger sponges in a $2\times2\times2$ configuration. This tiling produces a non-hollow central region while preserving the Hausdorff dimension of the original Menger sponge, despite breaking its global self-similarity. To see this, recall that the regular Menger sponge is generated recursively by retaining 20 smaller cubic copies at each iteration, each with a linear scaling factor of $1/3$. Its Hausdorff dimension is therefore given by $d_{\rm H}=\ln 20/\ln 3=\log_3 20\approx 2.727<3$. Since the tiled structure comprises a finite number of copies of the original Menger sponge, its Hausdorff dimension remains unchanged.

We begin with the cubic-lattice-regularized prototypical model of a 3D class AII topological insulator, namely, the 3D generalization of the BHZ model~\cite{TopoGen:8, TopoGen:12}, described by the Bloch Hamiltonian
\begin{equation}~\label{eq:ham_3d}
\begin{aligned}
H^{\rm BHZ}_{\rm 3D}(\mathbf{k})
&=t\left[\sin(k_xa)\Gamma_1+\sin(k_ya)\Gamma_2
+\sin(k_za)\Gamma_3\right] \\
&+\left[M-2B\sum_{\mu=x,y,z}
\left\{1-\cos(k_\mu a)\right\}\right]\Gamma_4.
\end{aligned}
\end{equation}
Here, $a$ denotes the lattice spacing, and the mutually anticommuting Hermitian $\Gamma$ matrices are defined as $\Gamma_1=\tau_3\otimes\sigma_1$, $\Gamma_2=\tau_3\otimes\sigma_2$, $\Gamma_3=\tau_3\otimes\sigma_3$, and $\Gamma_4=\tau_1\otimes\sigma_0$, where $\otimes$ denotes the tensor product. The two sets of Pauli matrices, $\{\tau_\mu\}$ and $\{\sigma_\mu\}$, act on the orbital and spin degrees of freedom, respectively. Depending on the ratio $M/B$, the model supports distinct topological insulating phases characterized by band inversions at different time-reversal-invariant momenta of the Brillouin zone. Specifically, for $0<M/B<4$ and $8<M/B<12$, band inversion occurs at the $\Gamma$ and ${\rm R}$ points, respectively, yielding strong topological insulators. In the intermediate regime $4<M/B<8$, simultaneous band inversions occur at three ${\rm X}$ and three ${\rm M}$ points, giving rise to a weak topological insulator. In contrast, the insulating phases for $M/B<0$ and $M/B>12$ are topologically trivial. The strong topological phases are characterized by a nontrivial $\mathbb{Z}_2$ invariant, protected by time-reversal symmetry and the associated Kramers' degeneracy of the bands. In what follows, we primarily focus on the strong topological phase with $0<M/B<4$, while extending some of our conclusions to the weak topological regime.

The tight-binding Hamiltonian corresponding to the Bloch Hamiltonian $H^{\rm BHZ}_{\rm 3D}(\mathbf{k})$ in Eq.~\eqref{eq:ham_3d}, defined on the parent cubic lattice, takes the form
\begin{eqnarray}
H^{\rm BHZ}_{\rm 3D, TB} &=& \sum_{\vec{r}} \bigg\{ \bigg( \sum_{\mu=x,y,z} \Psi^\dagger_{\vec{r}} \left[\frac{t}{2i} \Gamma_\mu + B \Gamma_4 \right] \Psi_{\vec{r}+\hat{e}_\mu} \nonumber \\
&+& {\rm H.c.} \bigg) + \Psi^\dagger_{\vec{r}} \left( M+6 B\right) \Gamma_4 \Psi_{\vec{r}} \bigg\} + \cdots,
\end{eqnarray}
where $\Psi^\top_{\vec{r}}=(c_{\vec{r},+,\uparrow}, c_{\vec{r},+,\downarrow}, c_{\vec{r},-,\uparrow}, c_{\vec{r},-,\downarrow})$ is a four-component fermionic spinor and $\Gamma_{x,y,z}\equiv\Gamma_{1,2,3}$. Here, $c_{\vec{r},\tau,\sigma}$ annihilates a fermion at site $\vec{r}=(x,y,z)$, with orbital parity eigenvalue $\tau=\pm$ and spin projection $\sigma=\uparrow,\downarrow$ along the $z$ direction. The ellipsis denotes additional hopping terms required to implement PBCs. Once the sites belonging to the tiled Menger sponge are identified [black sites in Fig.~\ref{fig:hypercubic_illustration}(c)], the parent Hamiltonian $H^{\rm BHZ}_{\rm 3D, TB}$ can be cast in the block-matrix form $H_{\rm parent}$, with the red sites representing its complement. Following the prescriptions outlined in Sec.~\ref{subsec:methods}, we then construct the tiled-Menger-sponge Hamiltonian $H_{\rm SL}$ using site elimination (method 1) and $H_{\rm R}$ using projection or renormalization (method 2). For the patterned or substituted-mass construction, we assign distinct on-site mass parameters, $M_0$ and $M_{\rm alt}$, to the black and red sites, respectively. Throughout our numerical calculations, we set $t=B=1$.

\begin{figure*}[t!]
    \centering
    \includegraphics[width=1.0\linewidth]{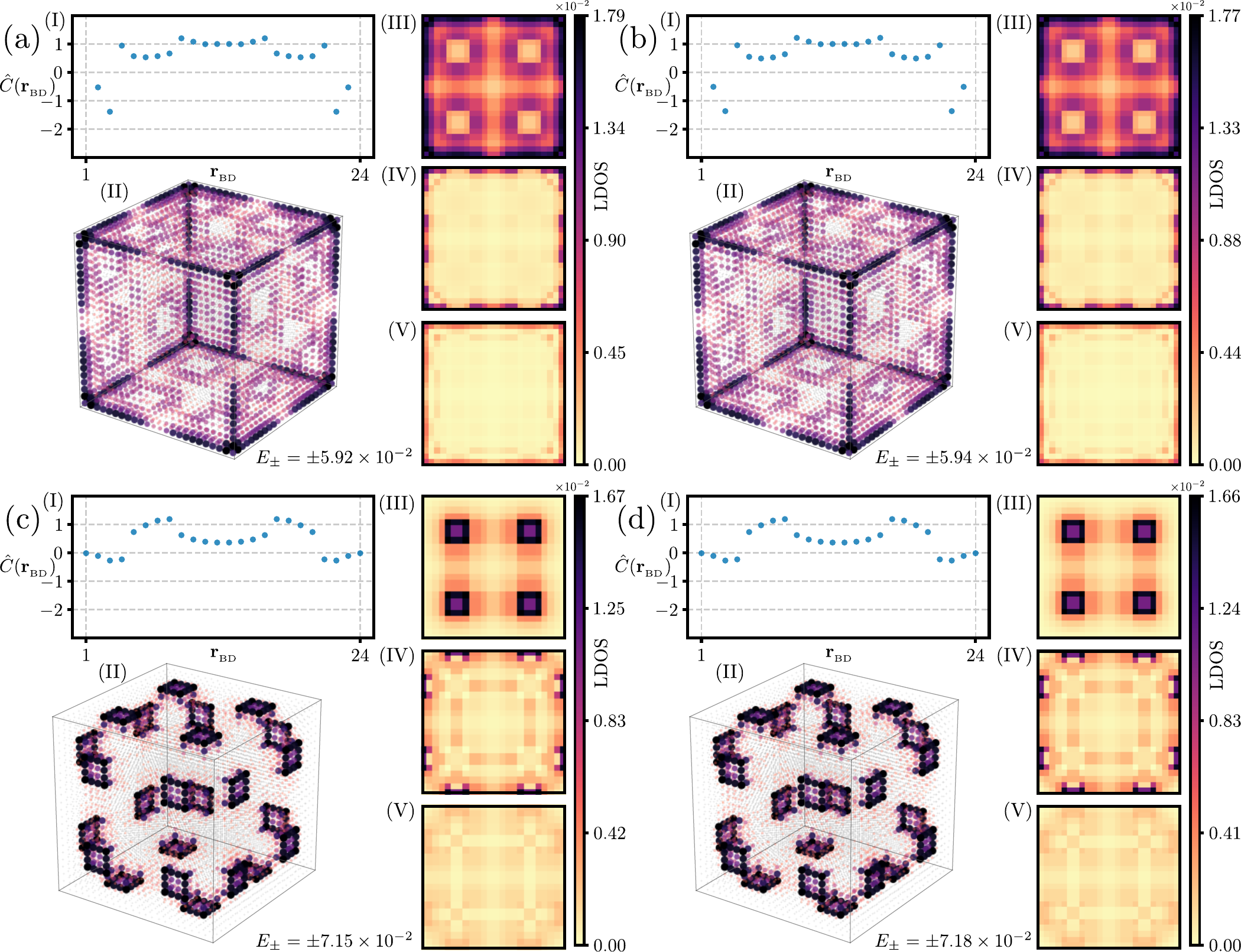}
    \caption{Local topological marker $C_{\rm 3D}(\vec{r})$ on the sites along the body diagonal $\vec{r}_{_{\rm 3D}}$ [(I)] and the local density of states (LDOS) associated with two particle-hole symmetric closest-to-zero-energy modes at energies $E_\pm$ (see legends) on a $L=24$ patterned fractal-mass tiled Menger sponge with open boundary conditions [(II)] for (a) $(M_0,M_{\rm alt})=(2.00,-0.05)$, (b) $(M_0,M_{\rm alt})=(2.00,-0.10)$, (c) $(M_0,M_{\rm alt})=(-0.05, 2.00)$, and (d) $(M_0,M_{\rm alt})=(-0.10, 2.00)$, where the largest 20\% LDOS values are projected onto the lattice while the remaining sites are masked to a grey color. The LDOS on the two-dimensional slices of the parent cubic system at (III) $z=1$, (IV) $z=7$, and (V) $z=13$.  
     }~\label{fig:sponge_substituted_marker}
\end{figure*}

Unlike in lower-dimensional systems, a global topological invariant for 3D class AII insulators cannot be readily obtained by diagonalizing the corresponding real-space tight-binding Hamiltonian. We therefore characterize their topology using a local topological marker, whose value at each lattice site is encoded in the corresponding diagonal matrix element of the operator~\cite{universalmarker}
\begin{equation}~\label{eq:utm_3d}
    {\mathbf C}_{\rm 3D}=N_D \; W \; \left[ {\mathbf Q} \; {\mathbf X} \; {\mathbf P} \; {\mathbf Y} \; {\mathbf Q} \; {\mathbf Z} \; {\mathbf P} + {\mathbf P} \; {\mathbf X} \; {\mathbf Q} \; {\mathbf Y} \; {\mathbf P} \; {\mathbf Z} \; {\mathbf Q} \right],
\end{equation}
where ${\mathbf X}$, ${\mathbf Y}$, and ${\mathbf Z}$ are diagonal position operators with matrix elements ${\rm X}_{ij}=x_i\delta_{ij}$, ${\rm Y}_{ij}=y_i\delta_{ij}$, and ${\rm Z}_{ij}=z_i\delta_{ij}$, respectively, and $\vec{r}_i=(x_i,y_i,z_i)$ denotes the position of the $i$th lattice site. The operator $W$ is defined as $W=\Gamma_5$, where $\Gamma_5=\Gamma_1\Gamma_2\Gamma_3\Gamma_4=-\tau_2\otimes\sigma_0$ is the fifth Hermitian matrix that anticommutes with the four $\Gamma$ matrices appearing in $H^{\rm BHZ}_{\rm 3D}(\vec{k})$ and $H^{\rm BHZ}_{\rm 3D, TB}$. The normalization constant is $N_D=1/(8\pi i)$. Since all systems considered here possess cubic symmetry, we evaluate and present the local topological marker along a body diagonal.

We begin by examining the spatial distribution of the local topological marker on the parent cubic lattice. For $M=-2$, the marker vanishes identically throughout the lattice, confirming the topologically trivial nature of the insulating phase. In contrast, for $M=2$ and $6$, it assumes quantized bulk values of $+1$ and $-2$, respectively, signaling distinct topological insulating phases. As expected for a local topological marker, its value also vanishes near the system boundaries, even in these topological phases. These results are presented in Fig.~\ref{fig:sponge_marker}(a). We obtain qualitatively similar results for the tiled Menger sponge, irrespective of whether its Hamiltonian is constructed using site elimination (method 1) [Fig.~\ref{fig:sponge_marker}(b)] or projection/renormalization (method 2) [Fig.~\ref{fig:sponge_marker}(c)].

We next establish the bulk-boundary correspondence by examining the gapless surface states supported by these systems under OBCs, thereby corroborating the topological characterization provided by the quantized bulk values of the local marker. Owing to the computational demands of exact diagonalization, which is required to evaluate the local topological marker, and our limited numerical resources, we restrict this analysis to the first-generation ($g=1$) tiled Menger sponge. Nevertheless, guided by our explicit results for 1D and 2D fractal lattices, we expect these conclusions to extend to higher generations ($g>1$) of the 3D tiled Menger sponge.

For $M=-2$, the LDOS associated with the two particle-hole-symmetric modes closest to zero energy is predominantly distributed throughout the bulk, without any discernible boundary localization, for both the parent cubic lattice [Fig.~\ref{fig:sponge_marker}(d)] and the tiled Menger sponge constructed using method 1 [Fig.~\ref{fig:sponge_marker}(e)] and method 2 [Fig.~\ref{fig:sponge_marker}(f)]. This bulk-like character is further corroborated by the layer-resolved LDOS evaluated on selected 2D planes perpendicular to the $z$ direction. In contrast, for $M=2$, the LDOS of these near-zero-energy modes becomes predominantly localized on the outermost surfaces, as clearly demonstrated by the corresponding layer-resolved profiles. This behavior is consistently observed for the parent cubic lattice [Fig.~\ref{fig:sponge_marker}(g)] and the tiled Menger sponge constructed using method 1 [Fig.~\ref{fig:sponge_marker}(h)] and method 2 [Fig.~\ref{fig:sponge_marker}(i)]. Furthermore, the LDOS on the tiled Menger sponge displays the underlying fractal structure, especially in the method of site elimination, showing considerable LDOS weight near the inner surfaces of the self-similar network. A qualitatively similar surface-localization pattern emerges for $M=6$ in the parent cubic lattice [Fig.~\ref{fig:sponge_marker}(j)] and the tiled Menger sponge obtained using site elimination [Fig.~\ref{fig:sponge_marker}(k)] and projection or renormalization [Fig.~\ref{fig:sponge_marker}(l)]. A subtle distinction, however, is that the spectral weight on the outermost surfaces is somewhat weaker for $M=6$ than for $M=2$, possibly reflecting the weak topological character of the insulating phase. Also, the LDOS structure does not show any clear visible fratcal strcture in this case.

Having established the existence of strong and weak $\mathbb{Z}_2$ topological insulating phases on the tiled Menger sponge, the geometric descendant of the cubic lattice, we finally investigate the emergence of topological phases in its patterned or substituted-mass counterpart. For concreteness, we focus on the trivial ($M<0$) and strong $\mathbb{Z}_2$ topological ($0<M<4$) insulating regimes of the parent $H^{\rm BHZ}_{\rm 3D}$ model, from which the on-site mass parameters $M_0$ and $M_{\rm alt}$ are selected. Specifically, we consider four representative parameter combinations: $(M_0,M_{\rm alt})=(2.00,-0.05)$, $(2.00,-0.10)$, $(-0.05,2.00)$, and $(-0.10,2.00)$. The corresponding results are presented in Figs.~\ref{fig:sponge_substituted_marker}(a)--\ref{fig:sponge_substituted_marker}(d), respectively.

For the first two parameter combinations, the local topological marker approaches the quantized value of $+1$ (within numerical error bars) over a substantial fraction of lattice sites along the body diagonal, signaling a topological insulating phase. Consistently, the LDOS associated with the two particle-hole-symmetric modes closest to zero energy exhibits pronounced localization over the entire outer surface of the system, as further confirmed by the corresponding layer-resolved profiles. In contrast, for the latter two combinations, the local topological marker generally deviates substantially from any nontrivial quantized value, except at a few sites near the interfaces between regions characterized by trivial $M_0$ and topological $M_{\rm alt}$. This behavior suggests that the resulting insulating phase is globally topologically trivial. Interestingly, despite the absence of a nontrivial bulk topological marker, the LDOS of the two near-zero-energy modes remains strongly localized on the outer surface, with pronounced spectral weight concentrated in regions characterized by on-site mass $M_{\rm alt}$, which belongs to the topological regime of the parent 3D BHZ model. This contrast highlights the crucial role of the local topological environment in determining the spatial distribution of near-zero-energy boundary modes. Taken together, these findings provide compelling evidence for the emergence of topological phases and boundary-localized modes in patterned or substituted-mass tiled Menger sponge lattices.

\section{Stability of fractal topological insulators against disorder}~\label{sec:disorder}

Thus far, we have demonstrated the successful realization of topological insulating phases on one-, two-, and three-dimensional fractal lattices using both site elimination and projection or renormalization, as well as on their patterned or substituted-mass counterparts in the clean limit. While these findings are encouraging, establishing the stability of fractal-based topological insulators against disorder is equally important, particularly in view of their potential practical applications, as real materials inevitably contain impurities and imperfections. However, the local topological markers exhibit appreciable spatial fluctuations even in the clean limit, despite assuming the expected quantized values over the majority of lattice sites. Consequently, they are not ideally suited for reliably identifying disorder-driven transitions between topological and normal insulating phases. Such an analysis requires a global topological invariant that assumes distinct quantized values in the two phases. Among the systems investigated in this work, only the two-dimensional models realizing QAHIs possess such a global invariant, namely, the Bott index. We therefore focus on these systems to investigate the robustness of fractal-based topological phases against disorder.

To investigate the robustness of fractal-based topological phases against disorder, we focus on the Haldane model and consider the effects of random charge impurities on both the parent honeycomb lattice and its geometric descendant, the hexaflake lattice. We quantify the stability of the topological phases by computing the disorder-averaged Bott index. On the parent honeycomb lattice, pointlike charge impurities are introduced at every lattice site, with their onsite potentials drawn independently from a uniform box distribution $[-W/2,W/2]$, where $W$ denotes the disorder strength.

\begin{figure}[t!]
    \centering
    \includegraphics[width=1.0\linewidth]{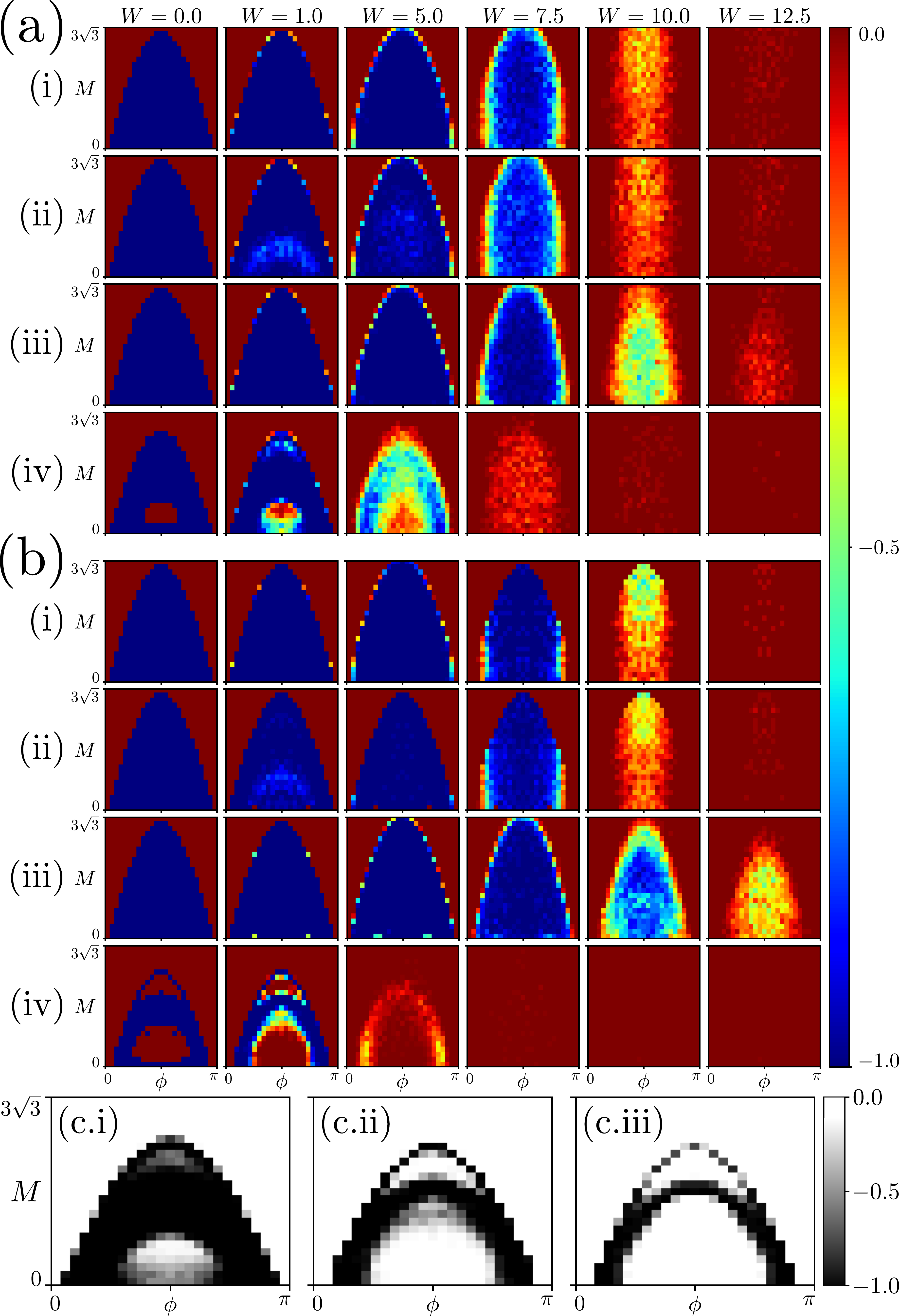}
    \caption{Phase diagram for the disordered Haldane model on honeycomb lattices [row (i)] with (a) 366 and (b) 3282 sites, in the presence of random charge impurities in terms of the disorder-averaged Bott index, also leading to similar outcomes in (a) second generation ($g=2$) and (b) third generation ($g=3$) hexaflake lattices [rows (ii)-(iv)] . On honeycomb lattice [(i)] pointlike charge impurities, drawn randomly from a box distribution $[-W/2,W/2]$ with $W$ denoting the strength of disorder, are added to each site, which is then projected to the hexaflake lattices [(ii)] to construct the respective phase diagram. In (iii), a clean honeycomb lattice-based Haldane model is projected onto the hexflake on which we subsequently add pointlike charge impurities to each of its sites. Row (iv) is obtained for a hexaflake, after the site elimination of the parent disordered honeycomb lattice Haldane model from (i). (c) Analogous to row (iv) of (a) and (b), showing the generation-dependence of the phase diagram for the hexaflake lattice with explicit results on (i) $g=2$, (ii) $g=3$, and (iii) $g=4$, with the asymptotic behavior shown in Fig.~\ref{fig:haldane_phase_diagram}(d). Each parameter value with finite $W$ is averaged over $50$ independent disorder realizations, and we show the results within the parameter region of $0 \leq M \leq 5.4$ and $0 \leq \phi \leq \pi$. Consult Sec.~\ref{sec:disorder}.
    }~\label{fig:haldane_disorder}
\end{figure}

Within the site-elimination approach, the disordered Hamiltonian on the hexaflake is constructed by removing, for each disorder realization, all sites of the parent honeycomb lattice that do not belong to the self-similar hexaflake network. In contrast, the projection or renormalization approach admits two distinct constructions. In the first, we project out the honeycomb-lattice sites lying outside the hexaflake in the presence of disorder. This procedure maps the onsite disorder of the parent lattice onto long-range-correlated disorder on the hexaflake, in addition to generating long-range hopping amplitudes. In the second construction, we first project the clean Haldane Hamiltonian onto the hexaflake and then introduce independent pointlike disorder potentials on its sites, again drawn from the uniform distribution $[-W/2,W/2]$. In this case, although the effective hopping amplitudes remain long-ranged, the disorder is short-ranged. These two distinct constructions lead to interesting results, which we discuss below.

The phase diagrams of the Haldane model on both the honeycomb and hexaflake lattices exhibit symmetries under $\phi\rightarrow-\phi$ and $M\rightarrow-M$. We therefore restrict our analysis to the parameter region $0\leq M\leq 5.4$ and $0\leq\phi\leq\pi$, in which the clean system has a Bott index of $-1$. For each parameter point in this region, we calculate the disorder-averaged Bott index by averaging over 50 independent disorder realizations. The results are presented in Fig.~\ref{fig:haldane_disorder}.

On the honeycomb lattice, the topological lobe remains robust against sufficiently weak disorder. As the disorder strength $W$ increases, the lobe gradually erodes from its boundaries, signaling the emergence of disorder-induced normal or trivial insulating phases. We find that a well-defined topological region of finite area, characterized by a disorder-averaged Bott index pinned at $-1$, persists up to $W=7.5$. Beyond this value, the system undergoes a smooth crossover toward a trivial insulating regime, where the disorder-averaged Bott index approaches zero, as exemplified by the results at $W=12.5$. Importantly, this crossover becomes increasingly sharp with increasing system size, suggesting a sharp topological phase transition in the thermodynamic limit. We present results for honeycomb lattices comprising 366 and 3282 sites in Figs.~\ref{fig:haldane_disorder}(a,i) and~\ref{fig:haldane_disorder}(b,i), respectively, whose geometric descendants are hexaflake lattices with generations $g=2$ and $g=3$. For the larger system, at $W=10.0$, we further identify a finite region in the $(\phi,M)$ parameter space where the disorder-averaged Bott index approaches the expected critical value of approximately $0.5$.

In the first renormalization scheme, where the disordered Haldane Hamiltonian is directly projected onto the hexaflake lattice, we find results that are quantitatively consistent with those obtained on the parent honeycomb lattice, irrespective of the system size or fractal generation. These results are presented in Figs.~\ref{fig:haldane_disorder}(a,ii) and \ref{fig:haldane_disorder}(b,ii) for $g=2$ and $g=3$, respectively. This agreement establishes the robustness of the Schur-complement projection procedure and its exactness even in the presence of disorder. In contrast, when random pointlike charge impurities are introduced only after projecting the clean Haldane Hamiltonian onto the hexaflake lattice, we observe enhanced robustness of the topological insulating phase against disorder. In particular, the topological lobe persists up to substantially larger disorder strengths, as shown in Figs.~\ref{fig:haldane_disorder}(a,iii) and \ref{fig:haldane_disorder}(b,iii) for $g=2$ and $g=3$, respectively. This enhanced stability can be understood from the correspondence between the clean parent and descendant systems: although the Haldane model on the honeycomb lattice is characterized by local hopping amplitudes and an on-site staggered mass, its hexaflake counterpart is governed by the long-range matrix elements of $H_{\rm R}$. Despite these differences in their microscopic descriptions, the two systems exhibit identical clean-limit phase diagrams due to exactness of Schur-complement projection procedure. Consequently, the longer-range couplings inherited by the hexaflake lattice provide additional robustness against random on-site disorder, requiring a larger disorder strength to destroy its topological phases. For a general discussion comparing these two approaches within the framework of method of projection or renoramlization, especially on the critical disorder strength, see Ref.~\onlinecite{GenConstruct:6}.

Finally, we investigate the stability of the QAHI emerging from the Haldane model on the hexaflake lattice within the site-elimination scheme in the presence of disorder. As in the clean limit, the sparse connectivity of the hexaflake lattice gives rise to a trivial insulating region embedded within the topological lobe, rendering the QAHI phase fragile. This feature persists in the presence of disorder as well, with the topological phase disappearing entirely for $5.0 \leq W \leq 7.5$, as shown in Figs.~\ref{fig:haldane_disorder}(a,iv) and \ref{fig:haldane_disorder}(b,iv) for $g=2$ and $g=3$, respectively. These results demonstrate that site elimination substantially compromises the robustness of the topological phase against disorder.

To further examine the stability of the QAHI in the thermodynamic limit, we fix the disorder strength at a moderate value of $W=1.0$ and compare the phase diagrams of hexaflake lattices with generation numbers $g=2$, $3$, and $4$, shown in Figs.~\ref{fig:haldane_disorder}(c,i), \ref{fig:haldane_disorder}(c,ii), and \ref{fig:haldane_disorder}(c,iii), respectively. We find that the fraction $p$ of the scanned parameter space occupied by the topological phase decreases monotonically with increasing $g$. However, when plotted against the inverse system size $1/N$, where $N$ denotes the number of lattice sites, $p$ appears to approach a finite asymptotic value ($p_{\rm sat}$) as $1/N \to 0$, as shown in Fig.~\ref{fig:haldane_phase_diagram}(e). Although the generation number $g$ and system size $N$ are intrinsically intertwined in the fractal geometry, which complicates a reliable extrapolation to the thermodynamic limit, the available data tentatively suggest that the QAHI phase may survive in the thermodynamic limit even under weak disorder within the site-elimination scheme.

\section{Summary and discussion}~\label{sec:summary}

In summary, starting from parent-lattice models of prototypical topological insulators in one, two, and three dimensions, belonging to different Altland–Zirnbauer symmetry classes, we demonstrate their realization on geometric descendant fractal lattices through two complementary approaches. We primarily focus on self-similar fractal geometries and construct the corresponding effective Hamiltonian ($H_{\rm SE}$) by either removing or eliminating the parent-lattice sites that lie outside the desired fractal network, an approach we term site elimination. Alternatively, we integrate out these same sites to obtain a renormalized Hamiltonian defined on the fractal lattice, an approach we refer to as projection or renormalization. Using these two complementary schemes, we successfully realize SSH topology on the Cantor set, Haldane-model QAHI phases on the hexaflake fractals, and strong $\mathbb{Z}_2$ and weak topological phases on the tiled Menger sponge fractal lattices. We characterize these phases using appropriate quantized local topological markers and global topological invariants, and establish their consistency with the bulk-boundary correspondence through the emergence of near-zero-energy modes localized at the outer and/or inner boundaries of the fractal lattices.

In the third approach, we retain the underlying lattice structure of the parent crystal but introduce a spatially patterned mass term, assigning distinct on-site mass parameters, $M_0$ and $M_{\rm alt}$, to the sites belonging to the target fractal network and its complement, respectively. These parameters are chosen from different topological regimes of the parent model, thereby generating what we term a patterned or substituted-mass fractal lattice. Using this approach, we demonstrate the realization of SSH topology on a patterned Cantor set embedded in a 1D chain, QAHI phases of the QWZ model on substituted-mass Sierpi\'nski carpets embedded in a parent square lattice, and strong $\mathbb{Z}_2$ topological insulating phases on patterned tiled Menger sponges embedded in a cubic lattice. In each case, the resulting topological phases are characterized by appropriate local topological markers or global topological invariants, with their bulk-boundary correspondence manifested through gapless modes localized at the outer and/or inner boundaries of the patterned fractal structures.

Finally, we demonstrate that QAHI phases on the hexaflake lattice remain robust against sufficiently weak potential disorder when constructed using either method 1 or method 2, irrespective of whether pointlike charge impurities are introduced before or after the projection in the latter approach. We anticipate this robustness to extend to other fractal-based topological insulating phases as well. Taken together, our findings suggest a general framework for realizing fractal-based topological insulators in arbitrary spatial dimensions, across different Altland–Zirnbauer symmetry classes, and starting from a broad range of parent crystals, both in the clean limit and in the presence of sufficiently weak disorder. These results open promising avenues for the experimental realization of topological phases on fractal lattices.

Designer electronic materials and optical lattices of ultracold atoms offer two promising quantum platforms for realizing topological insulators on fractal lattices. Highly tunable designer materials have already enabled the realization of regular lattices, quasicrystals, and fractal geometries~\cite{Gomes2012, Collins2017, frac:exp1, frac:exp2, frac:exp3, Ma2025, Yang2025}, providing versatile settings in which various topological models can be emulated. Similarly, desired fractal patterns can be engineered in optical lattices by selectively removing laser trap potentials. Several protocols for realizing different topological phases in optical lattices have recently been proposed~\cite{opticallattice:1, opticallattice:2, opticallattice:3, opticallattice:4}, which can be readily extended to their geometric descendant fractal networks. Among these, the experimental realizations of the Haldane model~\cite{opticallattice:3} and QWZ model~\cite{opticallattice:4} appear particularly promising. Of the three approaches proposed in this work for realizing crystal-based topological insulators on their geometric descendant fractal lattices, the substituted-mass patterning scheme may offer the most practical and experimentally accessible route for several reasons. First, it preserves the underlying lattice connectivity, thereby avoiding potential complications arising from structural modifications, such as unwanted band-bending effects that could compromise the stability of the topological phases and detection of boundary modes more challenging. Second, it requires only local control of the on-site mass parameter. In designer materials, this can be achieved by tuning the depth of the local trapping potential, while in optical lattices, it can be implemented through spatially selective modulation of the laser trap potential.

Beyond these promising quantum platforms, our theoretical predictions can also be tested using several highly tunable classical metamaterial platforms, among which topolectric circuits~\cite{Imhof2018, Dong2021, Olekhno2022}, photonic lattices~\cite{Ozawa2019}, and phononic or mechanical lattices~\cite{Susstrunk2015, Yang2015} are particularly promising. Various topological phases in different spatial dimensions have already been engineered in these systems, exhibiting robust boundary modes that remain resilient to inherent disorder. State-of-the-art experimental techniques in these platforms can be employed to implement the topological Hamiltonians associated with site-eliminated, projected, and patterned or substituted-mass fractal lattices. It is worth noting that some topological phases have already been realized on fractal metamaterials~\cite{frac:exp4, frac:exp5, frac:exp6, frac:exp7}. Such implementations can also incorporate the longer-range couplings intrinsic to the projection scheme (method 2), thereby providing a versatile route to systematically explore the rich landscape of fractal-based topological insulating phases across different embedding dimensions. We expect that the present theoretical work will trigger new experimental interests in this direction.

\acknowledgments 

This work was supported by NSF CAREER Grant No.~DMR-2238679 (USA) of B.R.

\end{document}